\PassOptionsToPackage{table,xcdraw}{xcolor}
\documentclass[twocolumn]{aastex631}

\usepackage[table,xcdraw]{xcolor}
\usepackage{changepage}

\usepackage{comment}

\usepackage{ulem}

\makeatletter
\newcommand*{\@rowstyle}{}
\newcommand*{\rowstyle}[1]{\gdef\@rowstyle{#1} \@rowstyle\ignorespaces}
\newcolumntype{=}{>{\gdef\@rowstyle{}}}
\newcolumntype{+}{>{\@rowstyle}}
\makeatother

\usepackage{amstext}
\usepackage{amsmath}
\usepackage{amssymb}
\usepackage{apjfonts}
\usepackage{graphicx}
\graphicspath{{./}{Figures/}}

\usepackage{savesym}
\savesymbol{tablenum}
\usepackage{siunitx}
\restoresymbol{SIX}{tablenum}

\usepackage{float}
\usepackage{placeins}
\usepackage{hyperref}

\begin{document}
\received{}
\revised{}
\accepted{}
\submitjournal{ApJ}

\shorttitle{R-process Mixing}
\shortauthors{Kolborg et al.}

\title{Constraints on the frequency and mass content of r-process events derived from turbulent mixing in galactic disks}

\correspondingauthor{Anne Noer Kolborg}
\email{anne.kolborg@nbi.ku.dk}

\author[0000-0001-7364-4964]{Anne Noer Kolborg}
\affiliation{Niels Bohr Institute, University of Copenhagen, Blegdamsvej 17, DK-2100 Copenhagen, Denmark}
\affiliation{Department of Astronomy \& Astrophysics, University of California, Santa Cruz, CA 95064, USA}

\author[0000-0003-2558-3102]{Enrico Ramirez-Ruiz}
\affiliation{Department of Astronomy \& Astrophysics, University of California, Santa Cruz, CA 95064, USA}
\affiliation{Niels Bohr Institute, University of Copenhagen, Blegdamsvej 17, DK-2100 Copenhagen, Denmark}

\author[0000-0001-9497-1374]{Davide Martizzi}
\affiliation{Department of Astronomy \& Astrophysics, University of California, Santa Cruz, CA 95064, USA}

\author[0000-0002-9946-4635]{Phil Macias}
\affiliation{Department of Astronomy \& Astrophysics, University of California, Santa Cruz, CA 95064, USA}

\author[0000-0001-7493-7419]{Melinda Soares-Furtado}
\altaffiliation{NASA Hubble Postdoctoral Fellow}
\affiliation{Department of Astronomy, University of Wisconsin-Madison, 475 N.~Charter Street, Madison, WI 53703, USA}

\begin{abstract} 
Metal-poor stars in the Milky Way (MW) halo display large star-to-star dispersion in their r-process abundance relative to lighter elements. This suggests a chemically diverse and unmixed interstellar medium (ISM) in the early Universe. This study aims to help shed light on the impact of turbulent mixing, driven by core collapse supernovae (cc-SNe), on the r-process abundance dispersal in galactic disks. To this end, we conduct a series of simulations of small-scale galaxy patches which resolve metal mixing mechanisms at parsec scales. Our set-up includes cc-SNe feedback and enrichment from r-process sources. We find that the relative rate of the r-process events to cc-SNe is directly imprinted on the shape of the r-process distribution in the ISM with more frequent events causing more centrally peaked distributions. We consider also the fraction of metals that is lost on galactic winds and find that cc-SNe are able to efficiently launch highly enriched winds, especially in smaller galaxy models. This result suggests that smaller systems, e.g. dwarf galaxies, may require higher levels of enrichment in order to achieve similar mean r-process abundances as MW-like progenitors systems. Finally, we are able to place novel constraints on the production rate of r-process elements in the MW, $6 \times 10^{-7} {M_\odot / \rm yr} \lesssim \dot{m}_{\rm rp} \ll 4.7 \times 10^{-4} {M_\odot / \rm yr} $, imposed by accurately reproducing the mean and dispersion of [Eu/Fe] in metal-poor stars. Our results are consistent with independent estimates from alternate methods and constitute a significant reduction in the permitted parameter space.
\end{abstract}

\section{Introduction}
The detailed physical ingredients required for r-process nucleosynthesis to take place were identified in the pioneering works of {\citet{1957RvMP...29..547B} and \citet{1957PASP...69..201C}. Despite that, the dominant astrophysical site for r-process production in the early Universe remains highly debatable \citep{Cowan2021}, even after the landmark discovery of the kilonova associated with GW170817 \citep{Abbott2017, 2017Sci...358.1556C,Kasen2017,Watson2019}. To this end, metal-poor stars in the galactic halo can be used as unique probes of r-process element synthesis in the early Universe and could help elucidate the astrophysical assets of the dominant progenitor system \citep{Sneden2008, Thielmann2017}.
 
Abundance similarities observed among metal-poor halo stars with ages discrepant by billions of years but with a distribution that is representative of the solar system \citep{Sneden2003, 1995AJ....109.2757M,Sneden2008, Roederer2014} suggest that nuclear pathways responsible for r-process elements are rather robust and have been operating coherently over extensive periods of time in the assembly history of the Milky Way (MW). From the large star-to-star chemical abundance dispersion in r-process elements (such as Eu), with respect to the $\alpha$ elements (such as Mg), we can infer that the injection of r-process elements occurs at a dramatically diminished rate when compared to core-collapse supernovae (cc-SNe)\citep{2000ApJ...529L..21W,2002ApJ...575..845F}. Additionally, the largest enhancements in [Eu/Fe] observed in metal poor stars imply that individual r-process sites need to synthesize a minimum of roughly $10^{-3}M_\odot$ r-process material \citep{2018ApJ...860...89M}. 

A central goal of this paper is to use a series of idealized hydrodynamic simulations of the response of galactic disks to SN feedback and metal mixing \citep{Kolborg2022} to provide a deeper physical interpretation of the star-to-star r-process scatter observed in the MW in the metallicity extent [Fe/H] of approximately -3.0 to -1.5, where Fe enrichment is driven primarily by cc-SNe \citep{Hotokezaka2018,2019ApJ...875..106C,Wanajo2021}. The constraints derived from the abundance pattern of r-process elements in the MW may ultimately help decipher the dominant production mechanism in the early Universe. 

The numerical modeling aimed at addressing the inhomogeneous enrichment of r-process elements in the MW has been primarily performed in a cosmological context \citep{Shen2015, VandeVoort2015,Naiman2018,Haynes2019, VandeVoort2020, VandeVoort2022}. These studies are limited by resolutions of a few tens to hundreds of parsecs, and they inevitably involve “sub-grid” conjectures for SN feedback, star formation and turbulent mixing. As such, large uncertainties remain at scales at which metal injection and turbulent metal mixing are taking place. In preference, in this paper we propose to model the gas interactions in small patches of a galaxy disk simulations in order to effectively isolate the small-scale mixing effects of metals at sub parsec scales from cc-SNe and r-process producing events (such as neutron star mergers (NSM)) into the interstellar medium (ISM). Besides being computationally workable \citep{Kolborg2022}, investigating the distribution of r-process elements in these simulations and comparing them extensively with observational data of metal poor stars can, in turn, help constrain the frequency of events and the mass content of r-process per event. 

This paper is structured as follows. In Section~\ref{sec:methods}, we summarize the simulation setups and introduce the key parameters relevant for metal production in a wide range of galaxy models. Sections~\ref{sec:UMW} through \ref{sec:prod_rate} present the results of our investigation. Initially, Section~\ref{sec:UMW} shows the results of the turbulent mixing study in a given galaxy potential. This section serves to introduce the reader to the salient concepts and build an understanding of the key mixing processes from which we build in subsequent sections. In Section~\ref{sec:potentials}, we present the mixing results across the three galaxy potentials studied, while in Section~\ref{sec:zNSM} we investigate how changing the allowed height of r-process events influences the mixing and ejection of r-process mass. Section~\ref{sec:prod_rate} investigates the production rate of r-process elements by studying how the spread of elements changes in response to variations in the mass per event (Section~\ref{subsec:mass-change}) and the relative rate of events (Section~\ref{subsec:rate-change}). This section concludes with a comparison between simulations and observations of [Eu/Fe] abundance of metal poor halo stars (Section~\ref{subsec:obs_match}). Section~\ref{sec:Discussion} gives a discussion of the implications of our simulation results, while Section~\ref{sec:conclusions} provides a final summary of our findings.

\section{Numerical Methods}
\label{sec:methods}
We use the hydrodynamic code RAMSES \citep{Teyssier2002} to simulate the evolution of gas in patches of galactic disks over timescales of hundreds of Myr. The fundamental simulation setup is the same as that used in \cite{Martizzi2016} and \cite{Kolborg2022}, with the inclusion of a passive scalar field in order to trace the enrichment of r-process elements. In this section, we describe the salient aspects of the models as they pertain to this project.
For additional details, we refer the interested reader to \cite{Martizzi2016} and \cite{Kolborg2022}.
The full suite of simulations used in this study are listed in Table~\ref{tab:sim_suite}. 

\begin{table*}[]
 \centering
 \begin{tabular}{ p {3 cm} | c | c | c | c | c | c | c | c | c | c | c }
 Galaxy model & $\Sigma_\mathrm{SFR}$ & $\rho_0$ & $z_\mathrm{eff}$ & $z_\mathrm{SNe}$ & $L$ & $dx$ & $T_{\rm end}$ & $\kappa$ & $z_\text{NSM}$ & $f_{\rm rp}$ & $m_{\rm rp}$ \\
 & M$_\odot$ / kpc$^2$ / Myr & g/cm$^3$ & pc & pc & pc & pc & Myr & pc km/s & $z_\text{SNe}$ & & $M_\odot$ \\ \hline
 MW progenitor, high SFR & \num{3e4} & \num{3.47 e-23} & 40 & 100 & 1000 & 3.9 & 120 & 147 & 1.33 & \num{1e-3} & \num{1e-2} \\
 & & & & & & & & & & & \num{1e-1} \\
 & & & & & & & & & & \num{1e-2} & \num{1e-2} \\
 & & & & & & & & & & \num{1e-2} & \num{1e-3} \\
 & & & & & & & & & 1.0 & & \num{1e-2} \\
 & & & & & & & & & 2.0 & & \\ \hline
 MW progenitor, low SFR & \num{1e3} & \num{2.08e-24} & 80 & 160 & 1000 & 3.9 & 250 & 213 & 1.33 & \num{1e-3} &\num{1e-2} \\ \hline
 Satellite & \num{8e2} & \num{4.67e-25} & 365 & 800 & 4000 & 15.6 & 500 & 1460 & 1.33 & \num{1e-3} & \num{1e-2} \\
 & & & & & & & & & & \num{1e-1} \\ 
 & & & & & & & & & & \num{1e-2} & \num{1e-2} \\
 & & & & & & & & & & \num{1e-2} & \num{1e-3}
 \end{tabular}
 \caption{Overview of all the simulations presented in this project. The columns are: name of the galaxy patch model, surface density of star formation rate, mid-plane density initial condition, effective scale height of the gaseous disk, maximum vertical height of a cc-SNe, box side length, cell size, total evolution time, turbulent diffusion coefficient \citep{Kolborg2022}, maximum vertical height of an r-process event (as a fraction of the maximum cc-SNe altitude), relative rate of r-process to cc-SNe events, and mass of r-process material per r-process event.}
 \label{tab:sim_suite}
\end{table*}

\subsection{Galaxy models}
The simulations assume a static gravitational potential produced by gas, stars, and dark matter as described by \cite{Kuijken1989}. The parameters of this gravitational potential are varied to emulate three different types of galaxies \citep{Kolborg2022}. 

The first model mimics an early MW progenitor with a high star formation rate (SFR). The second model simulates a MW progenitor with a more modest SFR. This second model is motivated by the recent findings of \cite{2021ApJ...915..116W}, who argue that the presence of the Large Magellanic Cloud satellite suggests a MW progenitor with a less active SFR. The third model mimics a weak gravitational potential with a high gas fraction and low SFR, emulating the expected properties of a classical dwarf galaxy. 

We employ cubic boxes\footnote{RAMSES does not yet allow tall box simulations} with periodic boundary conditions on the four edges perpendicular to the disk and outflow boundary conditions on the two parallel ones. The resolution of each simulation box is chosen such that the evolution of individual supernova remnants (SNR) are always well resolved. Specifically the cooling radius is resolved by at least 5 cells for least 94\% of all remnants, in all galaxy models \citep{Kolborg2022}. 

These various setups of galaxy disk patches at relatively high resolution, which capture the momentum injection of individual SNe, are able to effectively replicate the local conditions of galactic environments that might be representative of metal-poor stars assembled in the early MW and within accreted dwarf satellites.

\subsection{Core collapse supernovae and neutron star mergers}
cc-SNe and r-process events are seeded randomly with a constant rate (see the following section) and a flat distribution in space within fixed maximum allowed heights from the center of the galactic disk.

We model cc-SNe using the sub-grid model for SN feedback implemented by \cite{Martizzi2015}. In this model, each cc-SN event has an ejecta mass equal to the initial mass function (IMF) weighted average ($M_{\mathrm{ej}} = \SI{6.8}\,{M_\odot}$) and energy $E = \SI{1e51}\,\mathrm{erg}$. Metals are introduced into the gas by individual cc-SNe, which are assumed to be chemically identical. 

We apply the same sub-grid model of injection of mass and energy to the r-process producing events, however, we selected $M_{\rm ej} = \SI{1e-2}\,{M_\odot}$ and $E = \SI{1e51}\,\mathrm{erg}$. 
These values reflect the properties of the ejecta inferred in the gravitational-wave triggered NSM event GW170717 \citep{Kasen2017}. 

Although the morphology of a NSM remnant is highly asymmetric at early times \citep{2002MNRAS.336L...7R,2010ApJ...716.1028R,2011ApJ...736L..21R}, the subsequent radiative evolution is notably analogous to that of a SNR with similar total energy \citep{2016ApJ...830...12M}. The shell formation epoch, which occurs when the remnant becomes radiative, takes place by the time the mass of swept-up material reaches $M_{\rm c}\approx {10}^{3}{({n}_{{\rm{H}}}/1\,{\mathrm{cm}}^{-3})}^{-2/7}{M}_{\odot }$ \citep{2020ApJ...896...66K,2018ApJ...860...89M,Cioffi1988,Thornton1998,Martizzi2015}. 
The implications of this are twofold. First, sub-grid SN feedback models, like the ones used in this study, can be effectively used to resolve the key evolutionary phases of NSM remnants \citep{2016ApJ...830...12M}. 
Second, our r-process injection models can be applied whether enrichment has occurred via extremely rare cc-SNe \citep[e.g.,][]{2004PhT....57j..47C,2012ApJ...750L..22W,2015ApJ...810..109N,2018ApJ...864..171M,2019Natur.569..241S}, or through NSMs \citep[e.g.,][]{2010MNRAS.406.2650M,2011ApJ...736L..21R}, provided that any contribution of other freshly-synthesized metals (i.e., non r-process) is less than those contained in the swept-up ISM mass by the time the blast wave reaches the cooling phase \citep{Macias2019}. For the purposes of this study we focus our attention on NSMs as the source of r-process elements. 

%These two widely discussed r-process production mechanisms are expected to take place at different locations within the galaxy, with NSMs being displaced farther from their birth locations than cc-SNe. 
The maximum allowed height of standard cc-SNe events ($z_{\rm SNe}$) is fixed to twice the scale height of the gaseous disk. Fixing the allowed height of the cc-SNe to the gaseous scale height is motivated by the typical short distances traveled by massive stars between the time of their birth and the SNe. The r-process events, on the other hand, are allowed to occur within a region defined by the parameter $z_{\rm NSM}$. As a starting point, we set this value to $z_{\rm NSM} = 1.33 z_{\rm SNe}$. In Section~\ref{sec:zNSM}, we explore how changes to this scale height impact the turbulent mixing process, and we discuss the limitations of our galaxy patch simulations. 

\subsubsection{Event and metal injection rates}
The surface density rate of cc-SNe, $\Gamma$, is set by the rate of star formation in the galaxy patch, such that $\Gamma = \frac{\dot{\Sigma}_{\rm SFR}}{\SI{100}{M_\odot}}$. The star formation rate (and hence the cc-SNe event rate) is assumed to be constant. Due to the relatively short time span and the physical extent of the galaxy patches modelled, one can think of this as modelling a short star formation burst early in the history of the galaxy. \\
The rate of NSM events is set as a rate relative to standard cc-SNe, $\Gamma_{\rm NSM} = f_{\rm rp} \Gamma$. In this study we explore relative rates of $f_{\rm rp}$ in the range [$10^{-3},10^{-2}$], which is equivalent to 1 NSM event per [100,1000] cc-SNe. 

For the purposes of this project, we are interested in studying the spread of r-process elements relative to elements produced by cc-SN. We select Fe as our representative cc-SNe element, with each cc-SNe yielding $M_{\rm Fe} = \SI{8e-2}\,{M_\odot}$ \citep{Kolborg2022}. We model r-process events as NSMs and assume that the entire ejecta is predominately comprised of r-process elements with a total mass $m_{\rm rp} = M_{\rm ej, NSM}$. That is, r-process events are assumed not to contribute to Fe enrichment, in our simulations. 

As commonly carried out by the community, we designate Eu as the representative r-process element. We calculate the mass of Eu by applying the solar r-process abundance pattern \citep{Asplund2009,Sneden2008} and calculating the mass fraction of Eu to all the r-process elements. 

This is particularly useful when comparing with observations of metal-poor stars \citep{Sneden2008}. For the purpose of this study, we consider atomic mass number 69 as the strict lower limit for neutron capture element production ($A_{\rm min} = 69$). In Section~\ref{subsec:Hoto_comp}, we discuss the implications of this choice and the consequences for the results if $A_{\rm min} = 90$, which corresponds to the second and third r-process peaks \citep{Sneden2008}. 

$\alpha$, iron peak and r-process elements are tracked as individual passive scalar fields. For the purposes of this study only iron and r-process elements are of interest and the individual scalar fields allow us to make adjustments to $M_{\rm Fe}$ and $m_{\rm rp}$ in post-processing. This approach allows us to simulate a wide range of r-process masses per event, $m_{\rm rp} = [10^{-3},10^{-1}]{\,M_\odot}$, studying their effects on the metal enrichment of the ISM. \\

In \citet{Kolborg2022} we showed that the metals injected into the ISM by these cataclysmic events are mixed through the galactic disk by turbulent diffusion, which is in turn driven by the energy and momentum deposition from the most common events. The turbulent diffusion coefficient sets the timescale for metal mixing, which depends mainly on the scale height of the disk and the turbulent velocity dispersion \citep{Kolborg2022}. \citet{Martizzi2016} conducted resolution tests on the velocity dispersion in the MW progenitor with high SFR set-up and found that the turbulent diffusion coefficient is effectively converged at the resolution employed in this work. Furthermore, the turbulent driving scales in the MW progenitor models are $\gtrsim \SI{100}{pc}$ in both galaxy set-ups \citep{Martizzi2016}. This characteristic scale is nearly two orders of magnitudes greater than the resolution size scale of our models and, as a result, we expect the turbulent cascade to be well resolved. \\
The advantage of galaxy patch simulations is that they allow us to effectively capture the driving of the turbulence by cc-SNe, as well as, accurately studying its effects on the mixing of other freshly synthesized metals that are produced by much rarer cataclysmic events. The simulations are designed to isolate the mixing driven by cc-SNe feedback, which is primarily driven by turbulence and by galactic wind launching. By design, other important sources of mixing in galaxies, such as gas inflow, recycling of gas in galactic fountains, and disc shearing are not present in these simulations.

\subsection{Steady State}
All simulations are initialized with the gas in hydrostatic equilibrium with the static gravitational potential. As cooling and SNe feedback turn on simultaneously some of the thermal pressure support falls and turbulent pressure support increases. Hence, at early times, the disk is primarily supported by thermal pressure, while at later times, the disk is supported by both thermal and turbulent pressure driven by cc-SNe. 
Since information needs to be effectively communicated to all regions before turbulent motions in the bulk of the disk reach a statistical steady state, an initially transient phase is produced. The duration of this phase is usually well described by the characteristic relaxation time in our disk models, $t_r= 4z_{\rm eff} {\langle\sigma_v\rangle}^{-1}$, where $\langle \sigma_v \rangle$ is the time averaged, mass-weighted velocity dispersion of the gas \citep{Kolborg2022} .
In what follows, we neglect this initial phase, which lasts $\approx {1 \over 5} T_{\rm end}$ in all of our simulations (Table~\ref{tab:sim_suite}). In this work, we only discuss the results from the steady state of the simulations.

\section{Metal Mixing in high SFR environments}
\label{sec:UMW}
In this section, we examine the mixing of elements in the MW progenitor model with high SFR. 
For simplicity, we limit our attention to a single instance of this model where r-process events take place at a rate relative to cc-SNe of $f_{\rm rp} = 10^{-3}$ and each event yields a mass of r-process elements of $m_{\rm rp} = 10^{-2}\,M_\odot$. 
This assumes that cc-SNe occur in a region of $z_{\rm SNe} = \SI{100}{pc}$, while the r-process events occur in the region defined by $\vert z \vert \leq 1.33 z_{\rm SNe}$. 
We use this representative model to present a detailed account of the metal mixing features that may be present in the interstellar gas. These salient features can result naturally when one examines metal turbulent mixing by events occurring at disparate rates, yet their exact properties depend on the specific galaxy model and more sensitively on $f_{\rm rp}$ and $m_{\rm rp}$.

\subsection{The Tomography of a Galaxy Patch}
\begin{figure}
\centering
\includegraphics[width = 0.45\textwidth]{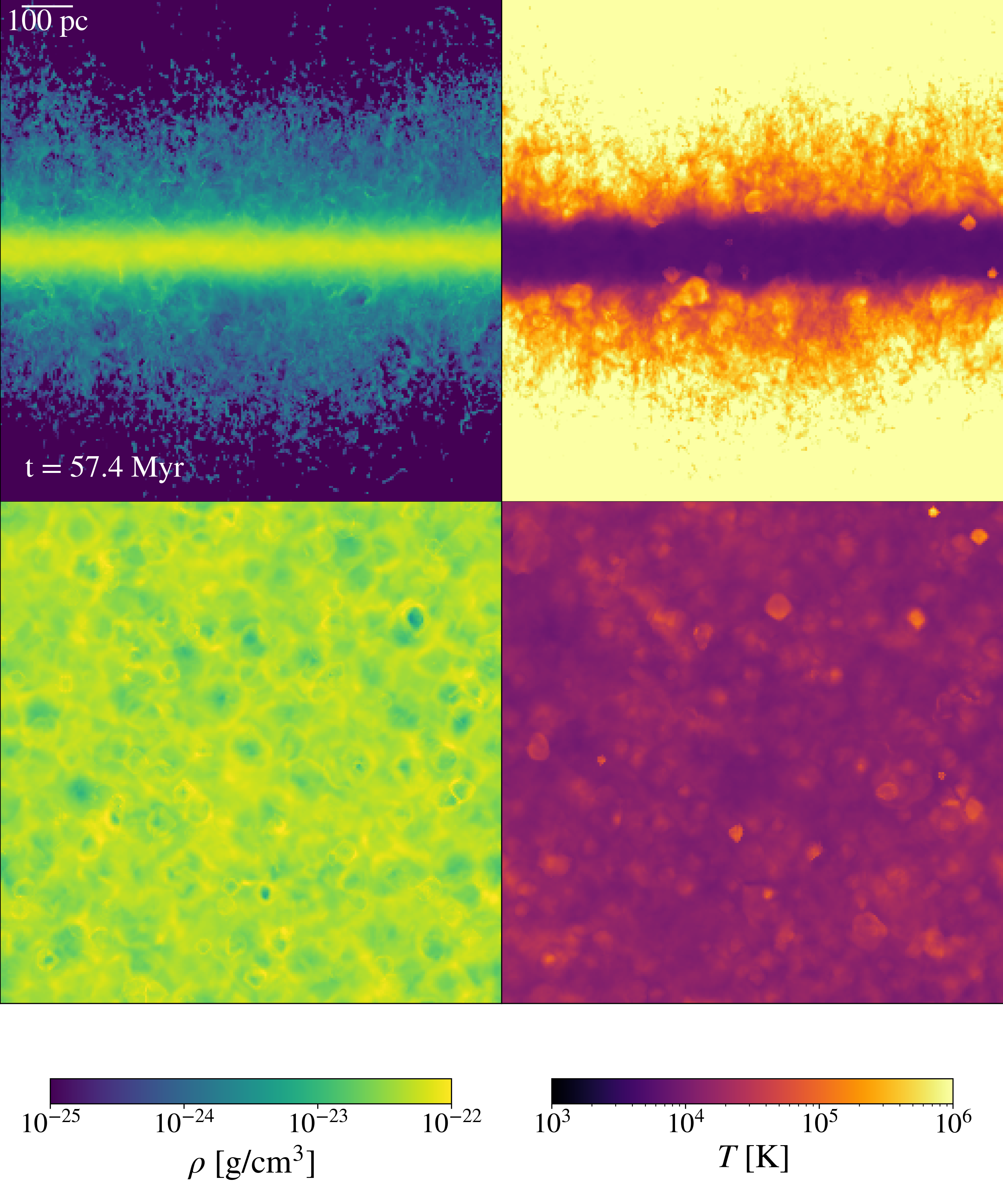}
 \caption{Density-weighted projection plots of the thermodynamic proprieties of the gas in the MW progenitor with high SFR (left-hand column: density; right-hand column: temperature). 
 The top row presents an edge-on view of the galaxy disk, while the bottom presents a face-on view. The SNe drive turbulent mixing in the disk and a galactic wind. Individual SNR are clearly visible in the plane of the disk.}
 \label{fig:UMW_rhoT_maps}
\end{figure}

\begin{figure*}
\centering
 \includegraphics[width = 0.95 \textwidth]{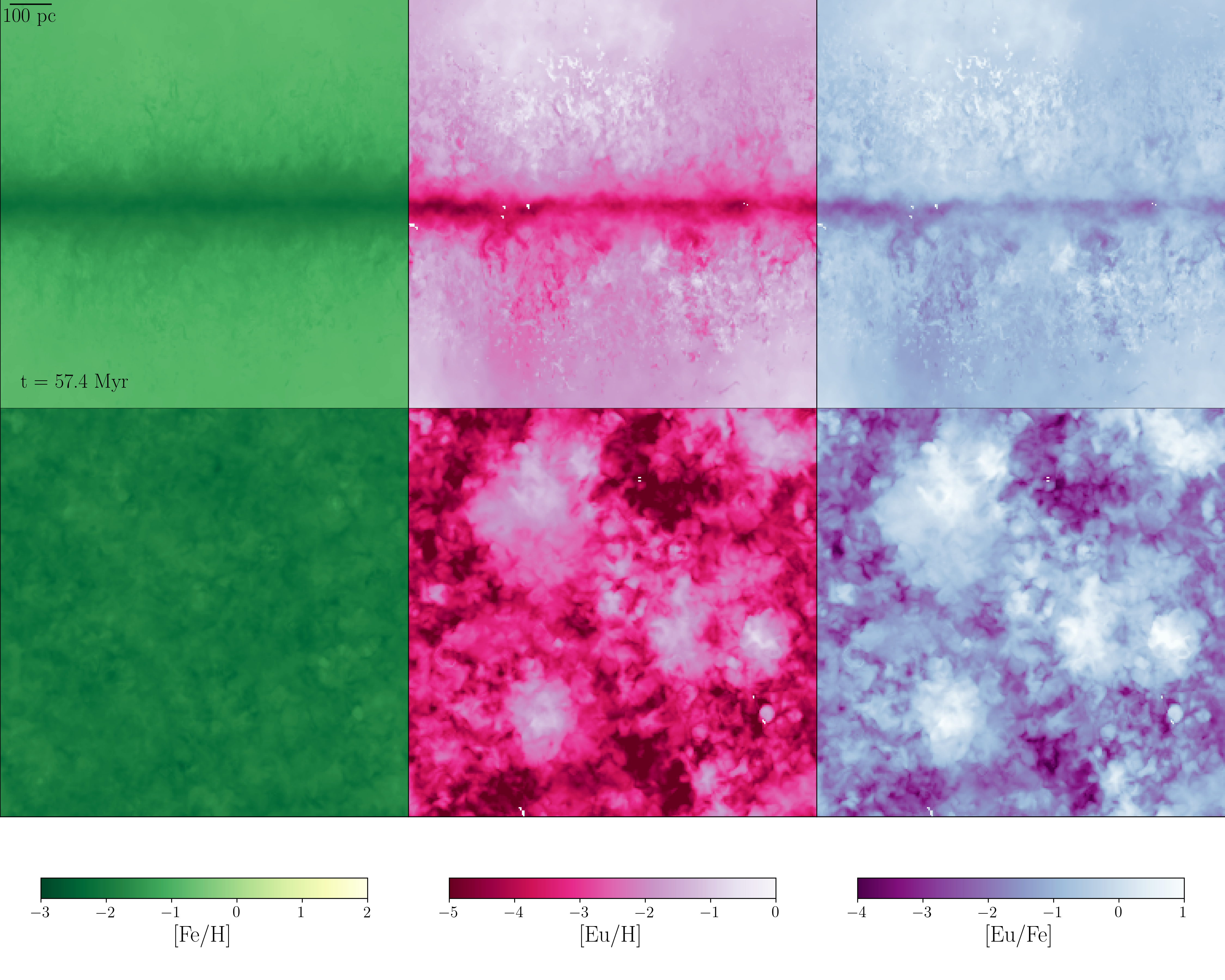}
 \caption{Density weighted projection plots of the metal mixing properties of the gas in the galaxy patch simulation shown in Figure \ref{fig:UMW_rhoT_maps}. The top row presents an edge-on view of the galaxy disk, while the bottom presents a face-on view.
 The panels show the abundances of [Fe/H] (left), [Eu/H] (middle) and [Eu/Fe] (right). Both Fe and Eu are efficiently ejected from the galactic disk and drastically increase the metal abundance in the galactic wind region. The effects of the relative rate of metal injection on the metal mixing of Fe and Eu are clearly seen; the less frequently injected Eu is less evenly distributed than Fe, which is injected more frequently.}
 \label{fig:UMW_metal_maps}
\end{figure*}

In Figure \ref{fig:UMW_rhoT_maps} we show the density-weighted projections of the thermodynamic properties of the gas along two axes in the simulation box. The density (left-hand column) and temperature (right-hand column) projections along an axis perpendicular to the galactic disk (top row) clearly illustrate how the momentum deposition from cc-SNe ejects gas from the disk in the form of a hot, rarefied galactic wind. From another point of view, the projection along the galactic disk plane (similar to a face-on view of the galaxy, bottom row) distinctively shows individual cc-SN and r-process remnants. They can be observed as nearly spherical regions of lower density and higher temperature gas, which are embedded in the denser, cooler, star forming gas.

In addition to momentum and energy, the individual events inject new metals: iron peak elements in the case of cc-SNe and heavy metals in the case of r-process events. These metals are mixed into the surrounding ISM by the turbulent diffusion, which is driven primarily by cc-SNe. 
Figure~\ref{fig:UMW_metal_maps} shows the density-weighted projections of the metal abundances in the gas for both Fe and Eu enrichment.

\begin{figure*}
 \centering
\includegraphics[width = 0.95 \textwidth]{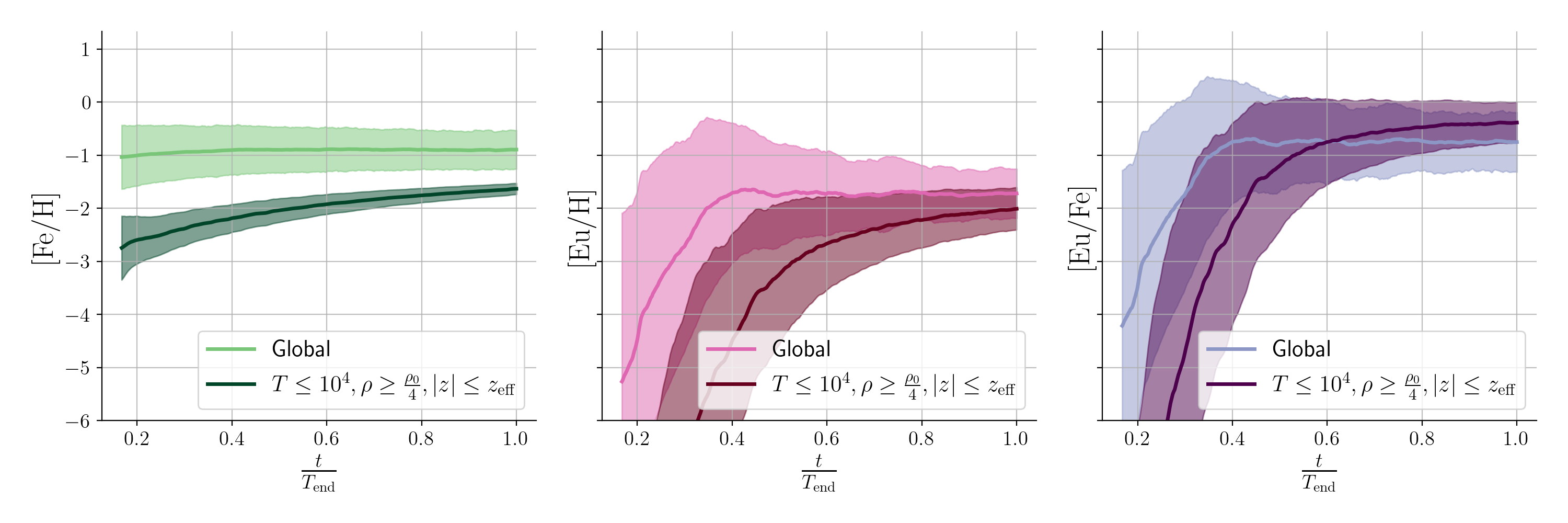}
 \caption{Temporal evolution of the volume weighted mean and $1 \sigma$ spread of the abundances of [Fe/H] (left), [Eu/H] (middle) and [Eu/Fe] (right) for the MW progenitor with high SFR. The global values refer to the mean and spread when considering all the gas in the simulation volume, while the cold, dense, disk material is defined by gas located at $\vert z \vert \leq z_{\rm eff}$ and with the following thermodynamic properties: $T \leq \SI{1e4}\,\mathrm{K}$ and $\rho \geq \rho_4 / 4 \approx \SI{1e-23}\,\mathrm{g\, cm^{-3}}$. The spread in [Fe/H] is always significantly less than the spread of [Eu/X] due to cc-SN metals being injected at much higher rates.}
 \label{fig:UMW_abundances}
\end{figure*}

Several key aspects can be noted by examining the maps shown in Figure~\ref{fig:UMW_metal_maps}. 
First, the rarefied galactic wind is highly enriched in Fe and Eu when compared to cold, dense gas in the mid-plane. This is primarily due to the low density of the hot wind within this region, which implies that even the mixing of a small amounts of metals can lead to a comparatively high enrichment. 

Second, the abundance of [Fe/H] is relatively uniform across the disk and within the galactic wind region. The full span of [Fe/H] variations is generally small. Across the disk, we observe localized [Fe/H] variations where neighboring cc-SN remnants remain compact and fail to overlap (at scales of a few tens of pc). 
These localized Fe enhancements are then subsequently smoothed out as the enriched gas is given more time to expand and mix throughout the disk \citep{Kolborg2022}. Yet, individual remnants always expand to characteristic lengths that are smaller than the disk scale height. 

Lastly, [Eu/H] is inspected in the middle panels of Figure~\ref{fig:UMW_metal_maps} to exhibit significantly larger scatter in abundance concentrations than [Fe/H] in the left panels. This originates from the comparative rarity of r-process metal injection, which naturally produces a chemically inhomogeneous and unmixed ISM at these early epochs. As more events are injected and r-process material diffuses and mixes, we can expect these metal concentrations to be gradually smoothed out.

\subsection{Metal mixing as a function of time}
\label{subsec:UMWabund}
From visual inspection of the abundance maps alone, we can gain a clear insight regarding the differential distribution of Eu and Fe in the galactic disk. In this section, we present a quantitative description of these differences. 

Figure~\ref{fig:UMW_abundances} depicts the volume weighted mean and $1\sigma$ spread of the gas abundances when considering all the gas in the simulation volume (global) and the cold, dense gas in the disk. For reference, the snapshots depicted in Figures~\ref{fig:UMW_rhoT_maps} and \ref{fig:UMW_metal_maps} correspond to a simulation time $57.4 {\rm Myr} \approx 0.475 T_{\rm end}$. 

Our scheme does not explicitly include star formation so, for the purpose of making an informed matching, we isolate the cold, dense gas within the disk ($\vert z \vert \leq z_{\rm eff}$), that is the most apt to form stars. We impose a temperature boundary of $T \leq 10^4\,\mathrm{K}$, which is related to the cooling function temperature floor, and a density limit of $\rho \gtrsim \rho_0/4$ (see Table~\ref{tab:sim_suite}).

As the simulation evolves, the [Fe/H] within the disk is gradually enriched as ensuing cc-SNe increase the mean Fe metallicity of the gas. At the same time, cc-SNe feedback drive outflows which cause Fe to spread from the disk onto the rarefied wind. The volume weighted average tends to emphasize the abundance of the hot, rarefied gas, which has a large volume filling factor. This becomes evident when examining the underlying distributions at fixed times shown in Appendix~\ref{app:distributions}. The global gas distribution of [Fe/H] has a very narrow peak at high metallicities (which drives the mean value) and a fairly long tail at lower metallicities (which influences the spread calculation). By contrast, the global and disk [Fe/H] distributions are rather similar when considering the mass weighted distributions. A similar behavior is seen for the [Eu/H] distributions, albeit with much broader "peaks" given the lesser degree of metal mixing which results from less frequent injections. 

Initially, the metal distributions of both [Fe/H] and [Eu/H] are highly inhomogeneous because there are significant regions that contain unmixed material. Due to its much more frequent injection, [Fe/H] is more uniformly distributed than [Eu/H]. The impact of higher metal injection rates on the abundance of [Fe/H] is twofold. 
First, the spread in the distribution is much narrower than the one seen for [Eu/H] in both the global and the cold, dense gas at all times in the simulation. In other words, there is a higher degree of homogenization of Fe than Eu as cc-SN products migrate and mix across the disk more effectively. 
Second, the evolution of the mean metallicity with time, which is caused by the rate of metal injection, is much swifter for [Fe/H] than for [Eu/H]. As time evolves, the localized inhomogeneities in both [Fe/H] and [Eu/H] are smoothed out and the evolution of the mean abundance becomes more gradual as each new injection of metals contributes progressively less to the total metal content. Yet r-process products migrate and mix throughout the box much more slowly. This is because large amount of freshly synthesized r-process material are injected at a rate that is usually faster than the the rate at which turbulent mixing smooths them down.

\subsection{Mass loading factor}
\label{sec:eta}
Through injection of energy and momentum, cc-SNe drive material out of the disk and launch galactic winds. In this section, we consider the mass loading factor of the wind, which is commonly defined as \citep[e.g.][]{Martizzi2016, Li2020}:
\begin{equation}
 \eta(z) = \frac{\dot{M}_{\mathrm{out}(z)}}{\mathrm{SFR}},
\end{equation}
$\dot{M}_\mathrm{out}$ denotes the rate at which mass is leaving the galaxy.

Analogously, the metal mass loading factor is defined as the ratio of the mass in metals leaving the galaxy to the ratio of mass of metals that are injected \citep{Li2020}:
\begin{align}
\eta_{Z_i}(z) &= \frac{\dot{M}_{Z_i, \mathrm{out}}(z)}{ \dot{M}_{Z_i, \mathrm{inj}}}.
\end{align}
Here we calculate the rate of metal injection as 
\begin{align}
\dot{M}_{Z_i, \mathrm{inj}} = \dot{n}_{\mathrm{SNe}} M_{\mathrm{ej}} y_{Z_i} f_{Z_i}.
\end{align}
$y_{Z_i}$ is the fractional yield of element $Z_i$ produced by a particular event ejecting a mass $M_{\rm ej}$, so that $M(Z_i)=M_{\rm ej} y_{Z_i}$. The parameter $f_{Z_i}$ denotes the fraction of events that produce metal $Z_i$ relative to the rate of cc-SNe.
For cc-SN elements $f_{Z_i} = 1.0$, while for r-process elements $f_{Z_i} = f_{\rm rp}$. 
 
The galaxy patch set-up is uniquely equipped to study the launching of galactic winds, which develop naturally from SNe feedback \citep{Martizzi2016} and require no sub-grid model. We estimated the rate at which mass leaves the disk by measuring the mass flux that streams through a surface parallel to the disk with an out-flowing z-velocity. We measure the outflow rate at three different heights. In order of increasing height from the disk midplane, the chosen distances are as follows: $\vert z \vert = 3 z_{\rm eff}$, $\vert z \vert = 2 z_{\rm SNe}$, and $\vert z \vert = \frac{L}{2} - dx$.The last measurement height is one cell size away from the edge of the simulation domain. The physical size of the relative heights are listed in Table~\ref{tab:sim_suite} for each galaxy model. 

It is important to note that the local boxes we consider in this study do not have a clearly defined escape velocity \citep{Martizzi2016} and the wind mass loading factor is observed to decline with increasing box height. \citet{Martizzi2016} studied SNe feedback in the galaxy patch simulations and argue that the structure of the galactic winds are not always well captured by these boxes. However, in the case of the MW progenitor with high SFR simulation (equivalent to their model FX-ULTRA-MW-L8), they found that the global wind properties are well modelled within $\vert z \vert \lesssim \SI{200}{pc}$, which is equivalent to $2 z_{\rm SNe}$ for that model. Therefore, we proceeded with the analysis of the wind loading, focusing on results from heights comparable to this value. Moreover, \citet{Li2020} compared loading factors from a wide range of simulations, including the work by \citet{Martizzi2016}, and found similar values reported across different global and local studies. This lends further credence to the robustness of the derived loading factors in this study.

Figure~\ref{fig:UMW_mass_loading} shows the evolution of the loading factors relating to the total mass, cc-SN elements, and r-process elements during the steady state of the simulation. The mass loading factor has a nearly constant value throughout the evolution of the simulation, with values that are consistent with the results of \cite{Martizzi2016}. The evolution of loading factor of Fe closely traces $\eta_{M}$. This seems natural to expect for a galactic wind driven primarily by momentum injection from the same cc-SNe that are also injecting Fe. Interestingly, $\eta_{Z_{\rm iron}}$, indicates that only about $\approx 10$\% of the injected metals are incorporated into the galactic winds. The evolution of $\eta_{Z_{\rm rp}}$, on the other hand, is much less smooth and contains large metal outbursts. The location of these outbursts closely follow the injection timing of r-process events, which are marked by dark vertical lines in Figure~\ref{fig:UMW_mass_loading}. 

The rise of r-process material in the wind takes place near the local injection sites, which produce large amounts of heavy elements. This freshly synthesized r-process material does not have time to mix effectively within the disk before being expelled. The smoothing of the individual outburst with $z$ is caused by r-process metals further mixing with wind material. It is noteworthy that the loading factors of cc-SN and r-process elements are very similar to each other at all times in the simulation, suggesting that r-process metals are carried out of the disk by winds driven by cc-SNe only after mixing significantly with the ISM. On average, the mass loading factor of iron elements is $\approx 7.5$\%, while the mass loading factor of r-process elements is $\approx 11$\% (both measured at $\vert z \vert = 2 z_{\rm SNe}$). The slightly higher mass loading factor implies that r-process elements are retained slightly less effective than cc-SN elements. This is closely related to the r-process metal outbursts captured in Figure~\ref{fig:UMW_metal_maps}, which occurs because individual r-process events do not have time to mix effectively with the surrounding ISM before being expelled from the disk. 

\begin{figure}
 \centering
 \includegraphics[width = 0.45\textwidth]{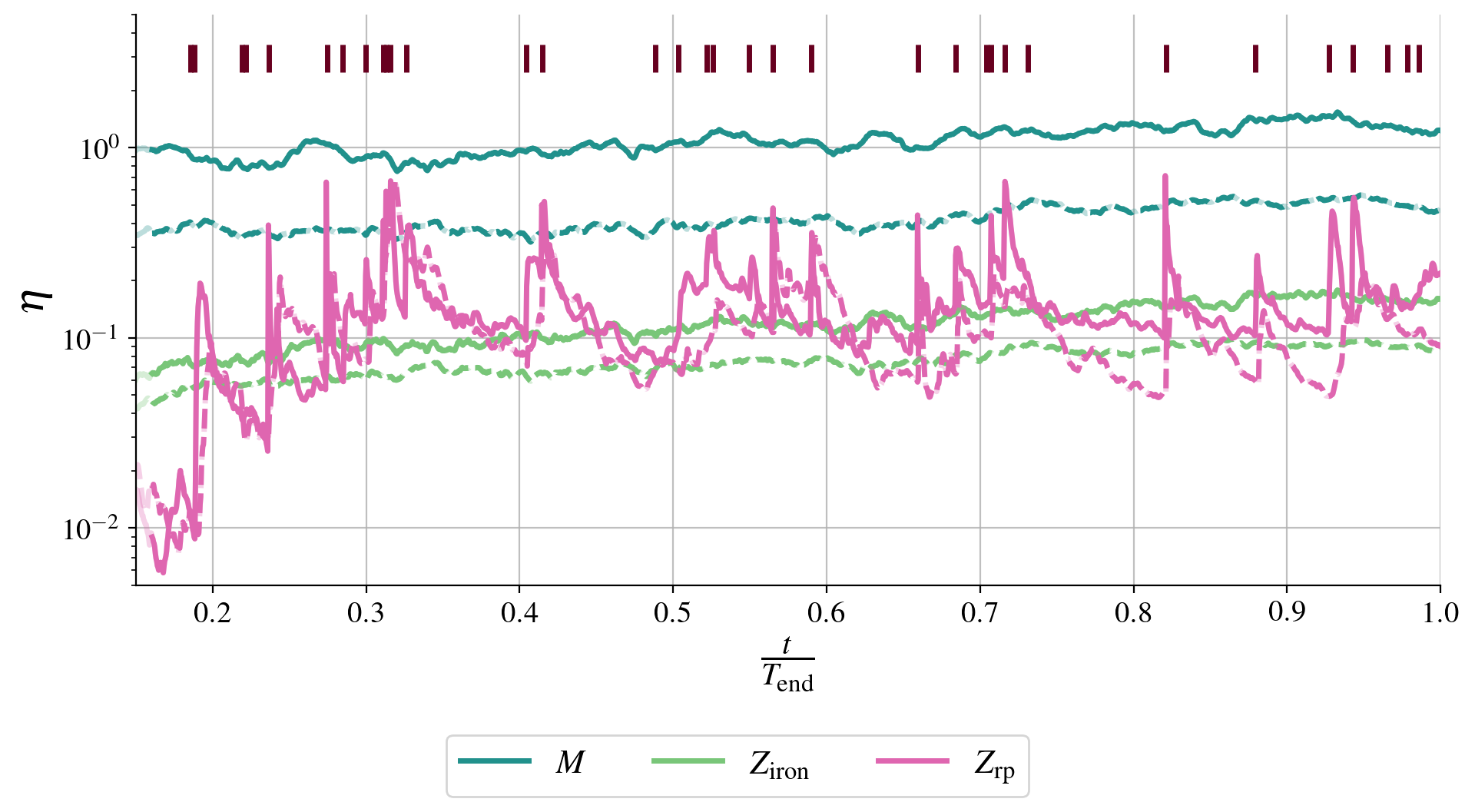}
 \caption{Evolution of the loading factors of the total, iron and r-process mass (see legend) as a function of time in the MW progenitor with high SFR. The loading factor is shown for two different heights in the potential: $\vert z \vert = 3 z_{\rm eff}$ (solid lines) and $ \vert z \vert = 2 z_{\rm SNe}$ (dashed lines). The dark vertical lines indicate the timing of r-process injections in the disk. The iron loading factor evolves very similarly to the total mass loading factor albeit with a different normalization. The r-process loading factor is strongly correlated with the injection of new r-process events. Interestingly, the loading factors of iron and r-process elements are comparable to each other, suggesting that both metal groups are carried away at similar rates irrespective of their very different injection rates. }
 \label{fig:UMW_mass_loading}
\end{figure}

\begin{figure*}
 \centering 
 \includegraphics[width = 0.95\textwidth]{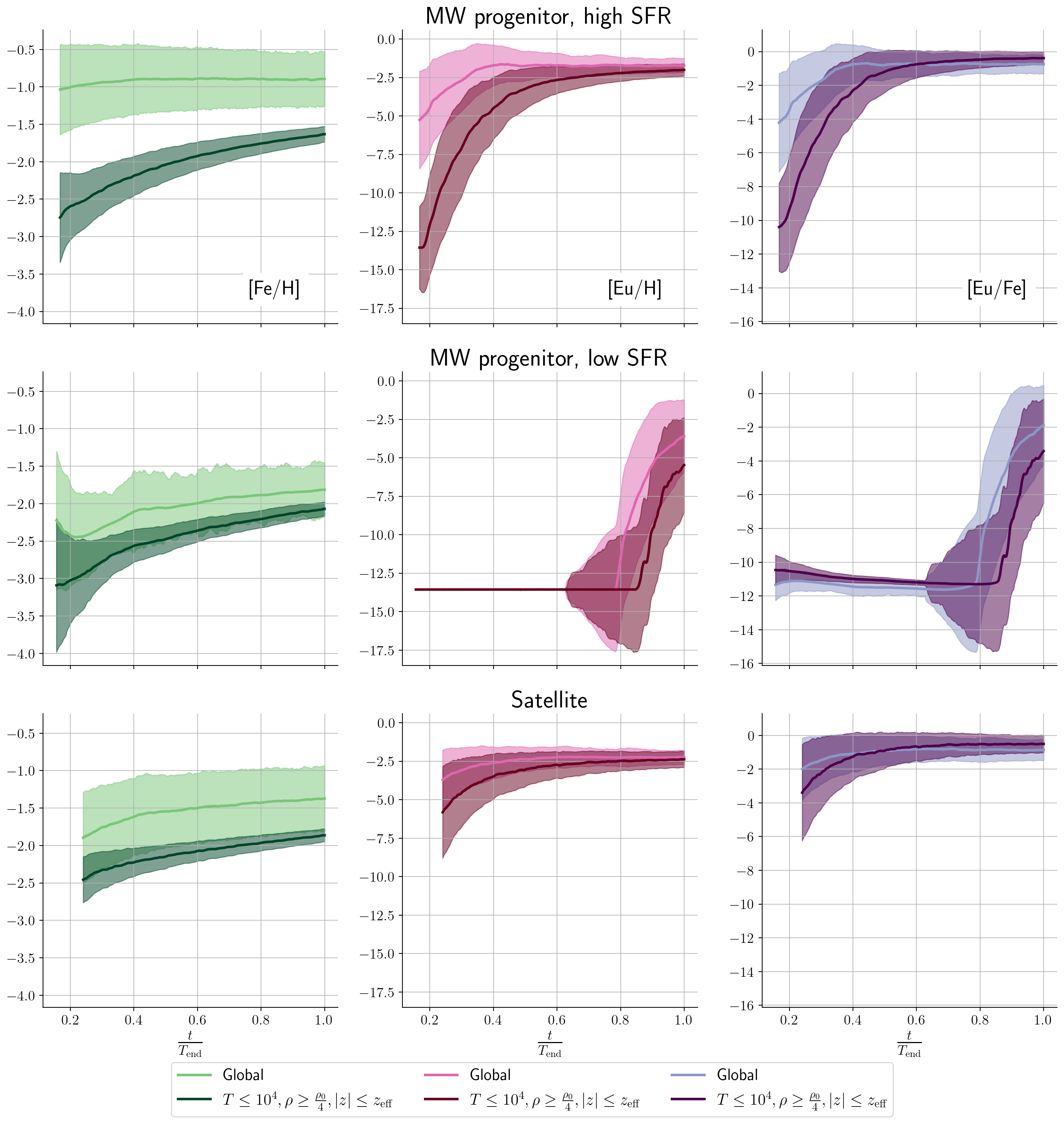}
 \caption{The temporal evolution of the volume weighted mean and $1 \sigma$ of [Fe/H] (left), [Eu/H] (middle), and [Eu/Fe] (right) in each of the three galaxy potentials.
 From top to bottom we present the results for the MW progenitor with high SFR, the MW progenitor with low SFR, and the satellite model. The abundances were calculated for all the gas in the box (global) and for the cold, dense gas in the disk as done in Figure~\ref{fig:UMW_abundances}. Overall, the evolution of the abundances are similar in all three galaxy potentials. Yet, the difference between the global and the cold, dense gas mean values is smaller for the weaker galactic potentials, implying larger volume filling factors of enriched gas in all gas phases. }
 \label{fig:all_abundances_large}
\end{figure*}

\section{Metal Mixing in different galaxy types}
\label{sec:potentials}
In this section, we consider the influence of the galaxy potential on the turbulent mixing of the freshly synthesized metals. To facilitate comparison, we assume standard values for all parameters that do not relate to the galaxy type.
More explicitly, the relative rate of NSM to cc-SN events is set to $f_{\rm rp} = 10^{-3}$, the mass per event is set to $m_{\rm rp} = 10^{-2}\,M_\odot$, and the scale height of NSM is set to $z_{\rm NSM} = 1.33 z_{\rm SNe}$. 

\begin{figure*}
 \centering
 \includegraphics[width = 0.95\textwidth]{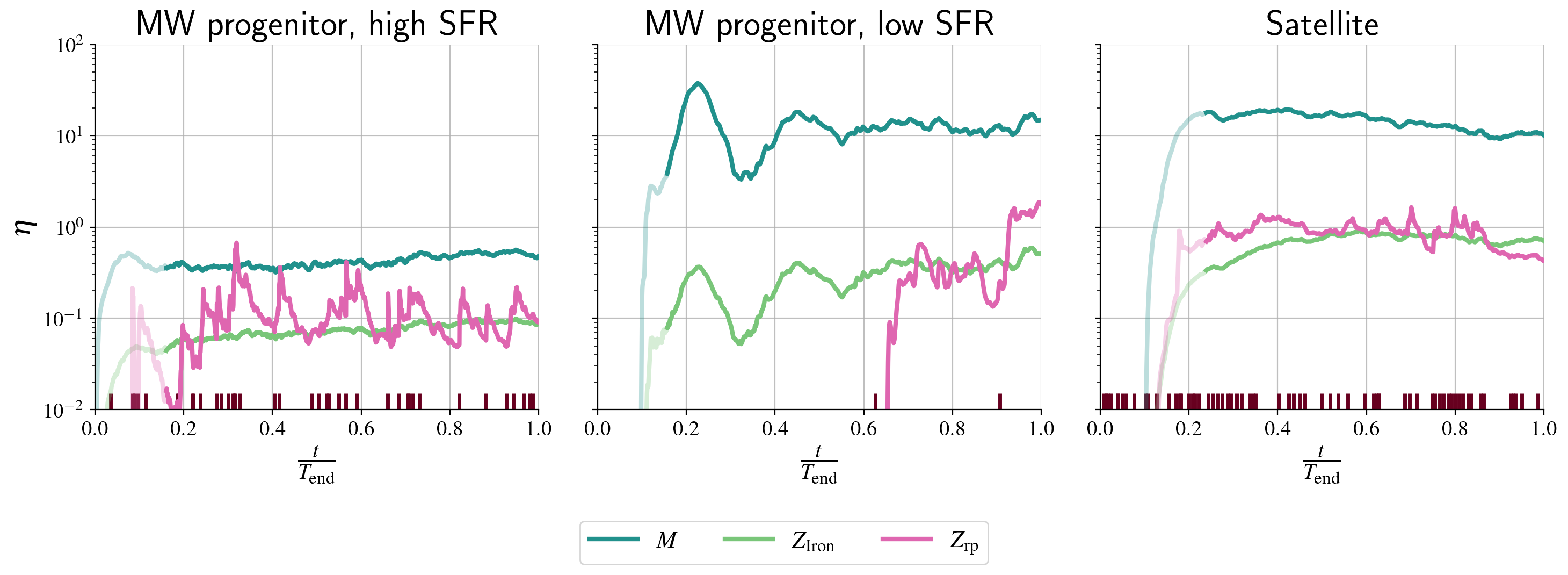}
 \caption{Loading factors of total, iron and r-process mass (see legend) measured at $\vert z \vert = 2 z_{\rm SNe}$ for three different galaxy potentials: MW progenitor with high SFR (left); MW progenitor with low SFR; and satellite galaxy (right). The dark vertical lines along the bottom of each panel denote the times at which r-process events take place. The times are normalized to the run times of the simulations (see Table~\ref{tab:sim_suite}). The initial phase (before steady-state is reached) is indicated by lower opacity color. The loading factors are generally larger for weaker gravitational potentials. The weaker gravitational potentials also display smaller differences in loading factors at greater distances from the disk (see Appendix~\ref{app:galaxy}), indicating the material streams from the disk more easily.}
 \label{fig:galaxy_mass_loading}
\end{figure*}

\subsection{Abundance evolution}
\label{subsec:all_abund}
Figure~\ref{fig:all_abundances_large} illustrates the evolution of the (volume-weighted) mean and spread of the [Fe/H], [Eu/H], and [Eu/Fe] abundances as a function of time in each of the three galaxy potentials. Similar to Figure~\ref{fig:UMW_abundances}, the mean and spread are shown for all the gas in the box (global) and the cold, dense gas within the disk. For all galaxy types, the global spread is observed to be larger than the spread of the cold, dense gas. We also see a significant offset between the mean abundance of the global gas and the cold, dense gas for all galaxy models. This offset is the smallest for the satellite model and is related to the lower gas density in the disk, which naturally leads to a higher volume filling factor of the enriched material in both the wind and the disk regions. 

As a whole, galaxy models with lower SFRs take longer to achieve the same mean [Fe/H] metallicity. As such, Fe is given more time to migrate throughout the disk. This is because the lower-density gas associated with a low SFR allows each cc-SN remnant to reach larger physical sizes than remnants in higher-density environments. As such, galactic disks with lower SFRs show less spread in abundances at a comparable average [Fe/H] metallicity. As anticipated, the spread of [Eu/H] is generally larger than that of [Fe/H], particularly in the cold, dense gas phase. 

In contrast with [Fe/H], the [Eu/H] spread is primarily driven by the relative rate of r-process to cc-SN events (which is the same in all simulations) and is less dependent on the SFR. It is compelling to note that the evolution of [Eu/H] for the MW progenitor with low SFR is rather odd (see middle panel in Figure~\ref{fig:all_abundances_large}). This transpires because the SFR is so low in this model that the galaxy patch is host to only two r-process events over its entire evolution. As a result, very metal poor stars in this model will not be enriched with any r-process material until after the first event takes place. This behavior clearly illustrates that galaxies with very low global SFRs might not be enriched with r-process material, as it is commonly expected to be the case for ultra faint dwarf (UFD) galaxies in the MW halo. Most of these systems are very metal poor and exhibiting low r-process enhancements \citep{Cowan2021}. A notable exception is Reticulum-II, which has been shown to contain several highly enriched stars \citep{Ji2016a,Ji2016b,Ji2022,Roederer2016}. In this context, it is interesting to consider the likelihood that any one small system hosts an r-process event given the very low SFRs associated with these systems.

\subsection{Mass loading factors}
One of our goals in this study is to understand the effects that turbulence driven by cc-SNe in galactic disks has on metal mixing. cc-SNe increase the mean metallicity of the gas in the disk and drive turbulent mixing, which prompts the metals to diffuse across the disk. At the same time, the momentum and energy that goes into driving turbulence launches galactic winds and drives metals out of the disk. 
Figure~\ref{fig:galaxy_mass_loading} shows the temporal evolution of the loading factors of mass, iron, and r-process metals for the three different galaxy potentials. The loading factors are all measured at $\vert z \vert = 2 z_{\rm SNe}$ (see Table~\ref{tab:sim_suite}). As expected, the mass loading factor is significantly larger for weaker galaxy potentials, highlighting the greater relative importance of cc-SN energetics in these systems. This result is consistent with what is frequently ascertained in dwarf galaxy simulations \citep[e.g.,][]{2017MNRAS.470L..39F}. 

During the simulations, the satellite galaxy model and the MW progenitor with low SFR release $\approx 35$\% and $\approx 10$\% of their initial mass to cc-SNe-driven winds, respectively. In contrast, the MW progenitor model with high SFR loses $\lesssim$5\% of its initial mass \citep{Kolborg2022}. This mass loss is naturally explained by the mass loading factors observed for these less massive systems. The associated larger mass loading factors in low mass systems are, for example, necessary in order to interpret the observed galaxy stellar mass function \citep[e.g.,][]{Li2020}.

Across the different galaxy potentials, the loading factor of iron and r-process elements are remarkably similar to each other, with a major caveat being that the MW progenitor with low SFR undergoes very low r-process mass loading before the first event occurs at $t \approx 0.6 T_{\rm end}$. 

Models with fewer cc-SNe give the freshly synthesized r-process metals more time to diffuse and travel across the disk and, as a result, show less prominent r-process outbursts when compare with the MW progenitor with high SFR. The average mass loading factor of iron and r-process elements in satellite galaxy model (MW progenitor with low SFR), at $\vert z \vert = 2 z_{\rm SNe}$, are $\approx 70$ \% ($\approx 30$ \%) and $\approx 89$\% ($\approx 22$ \%), indicating significantly less retention of metals in the ISM than the MW progenitor with high SFR. In essence, higher mass injection rates of r-process material are necessary in a satellite galaxy in order to average [Eu/H] abundance comparable to a more massive system. 

\section{Varying the scale heights of r-process injection and its effect on metal mixing}
\label{sec:zNSM}
Here we examine the impact of altering the scale height of r-process metal injection sites on metal mixing. 
The galaxy patch setup does not lend itself well to studying large offsets from the disk \citep[e.g.,][]{2003MNRAS.345.1077R,2007ApJ...665.1220Z,2010ApJ...725L..91K,2014ApJ...792..123B,2022ApJ...940L..18Z}, as would be expected for systems with long merger timescales ($t_{\rm delay}$) and large velocity kicks ($v_{\rm kick}$). This setup can, however, be used to probe the impact of modest event offsets, $\bar{z}\approx 140(t_{\rm delay}/10{\rm Myr})(v_{\rm kick}/20{\rm km\,s^{-1}})$ pc, on the subsequent mixing efficacy, as could be envisaged for r-process production in fast-merging double neutron stars \citep{2015ApJ...802L..22R,2019ApJ...872..105S} or rare cc-SNe \citep{2012ApJ...750L..22W,2015ApJ...810..109N,2018ApJ...864..171M}. These prompt channels are commonly argued to be more effective at enriching very metal poor stars with r-process products \citep[e.g.,][]{2019Natur.569..241S,Cowan2006}, which is the central focus of this study. 

In this section, we present results for three different simulations. For these runs, we use the same galaxy type. We select the MW progenitor with high SFR in view of the fact that, as we argued in Section~\ref{sec:potentials}, r-process products in this model are given the smallest amount of time to migrate throughout the disk. As such, the effects of varying the scale height of metal injection should be particularly obvious for this model. The relative rate and mass per event of r-process events remain unchanged. That is, $f_{\rm rp} = 10^{-3}$ and $m_{\rm rp} = 10^{-2}\,M_\odot$. Yet, the maximum allowed height of NSM is varied between the following three scale heights: $z_{\rm NSM} = z_{\rm SNe}$, $z_{\rm NSM} = 1.33 z_{\rm SNe}$, $z_{\rm NSM} = 2 z_{\rm SNe}$. To facilitate comparison, the injection procedure for cc-SNe remains the same in all simulations. 

\subsection{Mass loading of r-process elements}
It is natural to expect that the event scale height will influence the rate at which r-process material is lost via galactic winds. We explore this assumption here by examining the consequences of increasing $z_{\rm NSM}$ on $\eta_{Z_{\rm rp}}$. 

We expect $\eta_M$ and $\eta_{Z_{\rm iron}}$ to remain unchanged in these models as cc-SNe, whose injection properties are the same in all models, dominate the energy and momentum injection in the disk. In essence, cc-SNe are primarily responsible for driving both the turbulent mixing and the resulting galactic wind in these models. This is confirmed by our simulations which show that the average $\eta_M$ and $\eta_{Z_{\rm iron}}$ are similar to within a few percent in all simulations. 

In contrast, the average mass loading factor of r-process elements vary perceptibly with $z_{\rm NSM}$ (see Figure~\ref{fig:rp_sh}). More precisely, the values of $\eta_{Z_{\rm rp}}$ measured at $\vert z \vert = 2 z_{\rm SNe}$, are $\approx 8.2$ \% ($z_{\rm NSM} = z_{\rm SNe}$), $\approx 12.3$ \% ($z_{\rm NSM} = 1.33 z_{\rm SNe}$)\footnote{There is a small difference between the average reported here and in section \ref{sec:eta}. The number reported in this section includes only the time that is common among the three simulations, this shorter time span leads to small differences in the reported figures.} and $\approx 12.6$ \% ($z_{\rm NSM} = 2 z_{\rm SNe}$). As expected, the model where r-process events are confined to a region that is significantly larger than the one for cc-SNe shows the largest $\eta_{Z_{\rm rp}}$, indicating slightly less effective retention of r-process metals in the ISM. 

%Old values were for the maximum height and the percentages were: 4.6, 5.3 and 8.3, respectively

\begin{figure}
 \centering
 \includegraphics[width = 0.43 \textwidth]{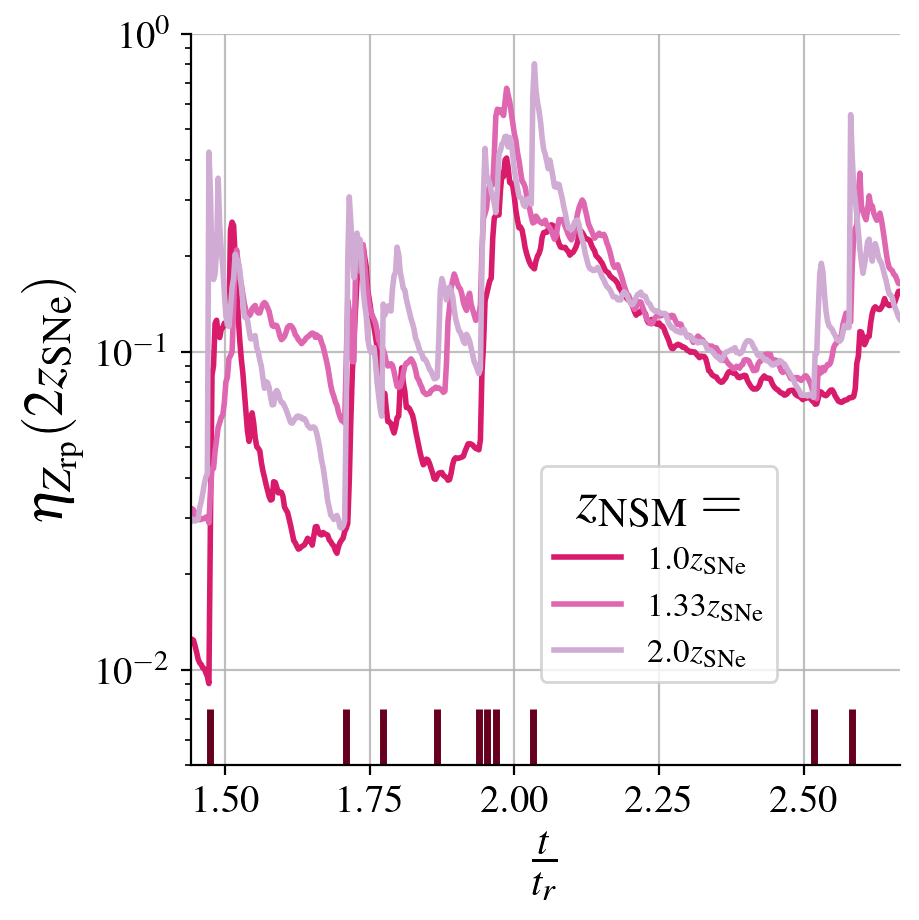}
 \caption{The temporal evolution of the metal r-process loading factor, $\eta_{Z_{\rm rp}}$, in units of the characteristic relaxation time, $t_r$, for varying $z_\mathrm{NSM}$ in the MW progenitor high SFR model. $\eta_{Z_{\rm rp}}$ is measured here at $\vert z \vert = 2 z_{\rm SNe}$. The dark vertical lines denote the times at which r-process events transpire. The r-process retention is slightly altered by changes in $z_\mathrm{NSM}$, with more prominent metal outbursts observed when injection takes place at higher scale heights.}
 \label{fig:rp_sh}
\end{figure}

\subsection{Abundance evolution}
\begin{figure}
 \includegraphics[width = 0.5 \textwidth]{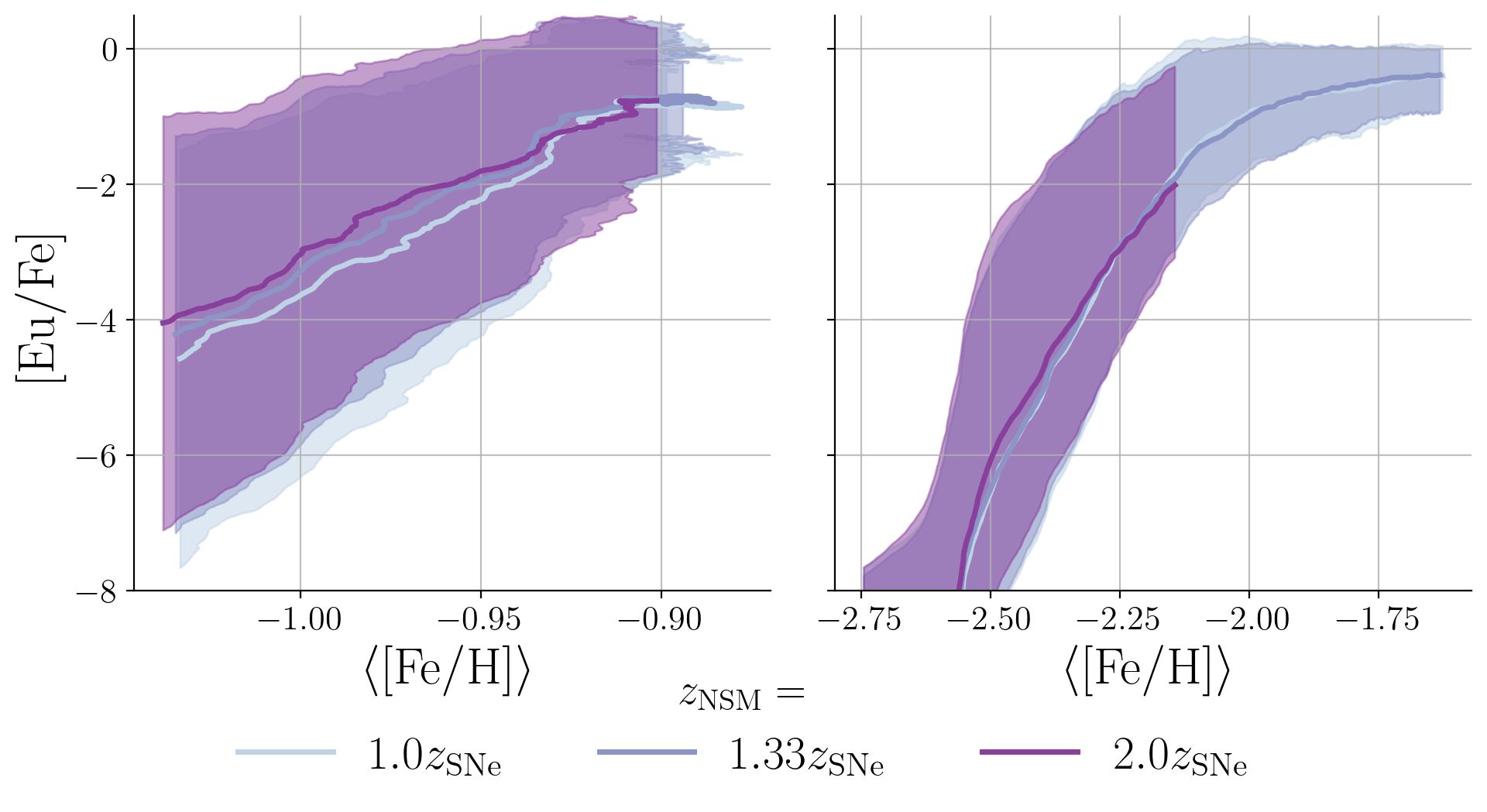}
 \caption{The temporal evolution of the volume weighted mean and $1 \sigma$ of [Eu/Fe] as a function of the global mean [Fe/H] abundance in MW progenitor with high SFR simulations with varying $z_\mathrm{NSM}$ (see legend). The [Eu/Fe] abundance is shown in the left panel for the entire simulation box and in the right panel for the cold, dense gas in the disk (defined as for Figures~\ref{fig:UMW_abundances} and \ref{fig:all_abundances_large}). We note no discernible difference in the mean abundance of the gas, suggesting that the offset of events does not strongly influence the observed abundances of r-process elements within the limits examined here.}
 \label{fig:UMW_event_height}
\end{figure}

Having found only a small increase in $\eta_{Z_{\rm rp}}$ with r-process event offsets, we now consider if the location of events are capable of altering the metal abundance of the ISM. 
Figure~\ref{fig:UMW_event_height} shows the volume-weighted mean and $1 \sigma$ of [Eu/Fe] as a function of the mean [Fe/H] for the global (left) and cold, dense, disk gas (right) in the three simulations. The definition of cold, dense disk gas remains unchanged from Sections~\ref{subsec:UMWabund} and \ref{subsec:all_abund}.

While there is a trend of slightly higher mean [Eu/Fe] abundances with increasing event offsets in the global measurement, we note no discernible difference when considering only the cold, dense gas. This is to be expected as the retention of r-process material in these models is $\gtrsim$ 90\%. Both the mass loading factors and abundances are found to be similar across models with different allowable scale heights of r-process events. Thus, we conclude that the range of offsets examined here do not produce measurable changes in observed galaxy properties. 

We caution the reader that global disk galaxy simulations are necessary in order to study the mixing effects of r-process metals deposited by NSM with large kicks and long merger times. This will require a large range of scales to be resolved simultaneously, which is exceedingly challenging given the complexity of the interplay between the various galaxy assembly mechanisms at all scales. Considering the somewhat emerging stage of modeling in the field, this study amounts to a sizable improvement within the stated limitations in our understanding of how the location of metal injection within the galaxy influences the inhomogeneous enrichment of the ISM and the mixing of r-process elements.
 
\section{On the repercussions of the production rate of r-process material on metal mixing}
\label{sec:prod_rate}
The purpose of this paper is to help isolate some of the key mechanisms that regulate turbulent mixing of r-process elements in galactic disks and, in particular, how the statistics of abundance variations in stars can help constrain astrophysical r-process synthesis models. It seems reasonable to expect that most astrophysical r-process production models would have r-process mass per event, $m_{\rm rp}$, and (relative) rate of events, $f_{\rm rp}$, as essential parameters. While a given r-process production mechanism may display a correlation or even interdependence between these two parameters, it is instructive to consider these two independently for understanding the mixing properties. Motivated by this, in this section, we examine more closely how the shape of the distribution of [Eu/Fe] abundances in the ISM changes in response to shifts in $m_{\rm rp}$ and $f_{\rm rp}$.

First, we consider changes in $m_{\rm rp}$, keeping all other parameters fixed. Next, we consider changes in $f_{\rm rp}$ while keeping $m_{\rm rp}$ constant. Finally, we address how the statistics of variations of metal poor halo stars in chemical space can provide valuable constraints on $m_{\rm rp}$ and $f_{\rm rp}$. Throughout this analysis, we focus on the results derived from the MW progenitor with high SFR model and the satellite galaxy model. 

\subsection{The role of $m_{\rm rp}$}
\label{subsec:mass-change}
The key to using galaxy patch models productively is to isolate the role of key parameters and then to analyse the simulations so that we can learn the role of these model ingredients on metal mixing. In this analysis we first begin by isolating the role of $m_{\rm rp}$. 

Figure~\ref{fig:mass_change} shows the stacked one-dimensional distributions of [Eu/Fe] within the cold, dense gas at each average [Fe/H] metallicity. The color bar notes the fractional mass in each bin. The filled-in contours represent the standard enrichment model (i.e., $f_{\rm rp} = 10^{-3}$ and $m_{\rm rp} = 10^{-2}\,M_\odot$), while the line contours denote the model with $m_{\rm rp}$ increased by a factor of 10 (i.e., $m_{\rm rp} = 10^{-1}\,M_\odot$). The left-hand panel shows our results from the MW progenitor with high SFR model, while the right-hand panel reveals our outcomes from the satellite galaxy model. 

In both galaxy models, as expected, the shape of the contours is unchanged by altering $m_{\rm rp}$. The main difference is an increase in the mean abundance of [Eu/Fe], as the mass of r-process injected over time is augmented in the box. The metal dispersion, on the other hand, stays the same as the relative rate of metal mixing, which is controlled by the cc-SN rate, remains unchanged. 

\begin{figure*}
 \centering
 \includegraphics[width = \textwidth]{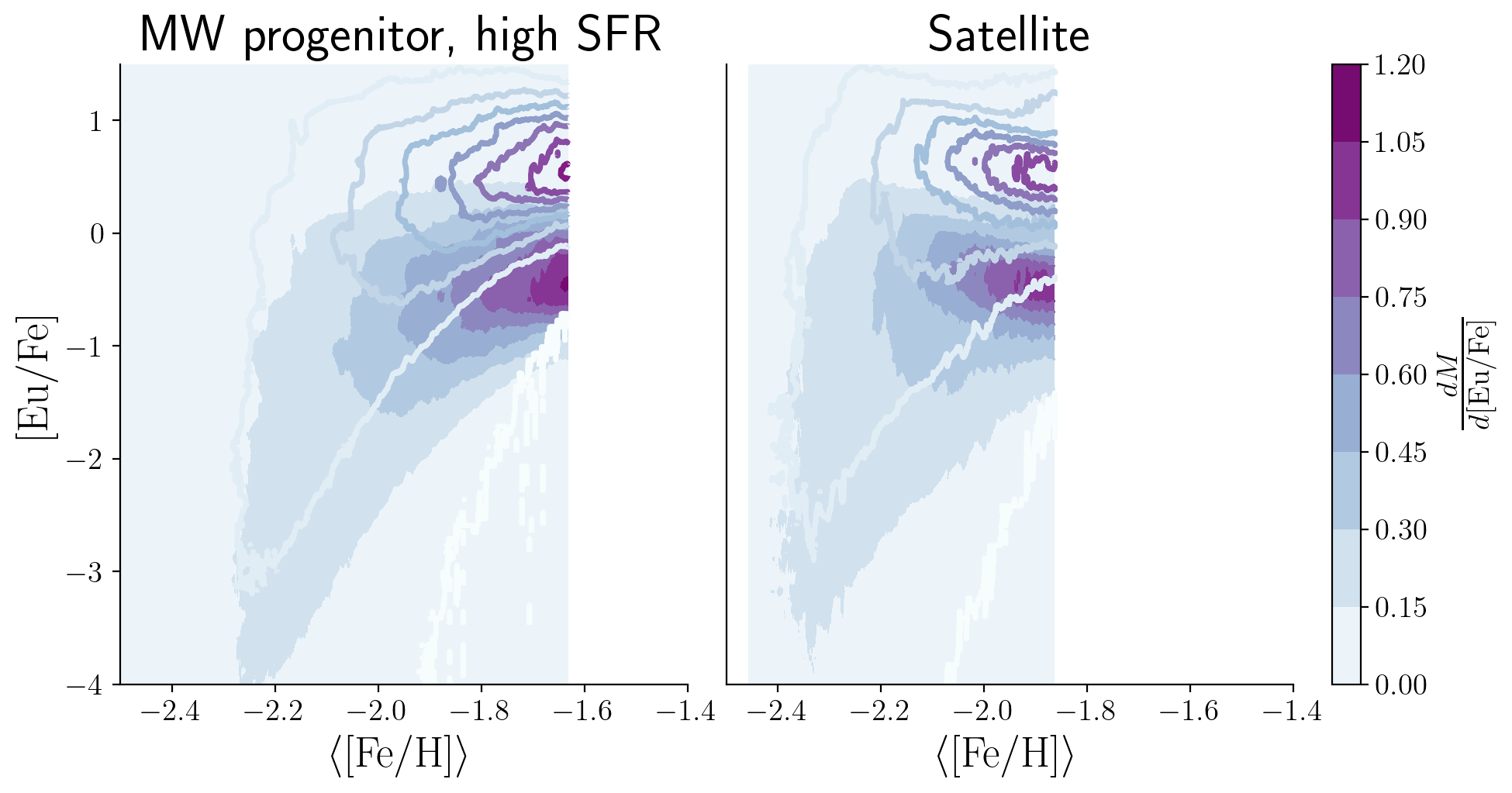}
 \caption{Stacked 1d distributions of [Eu/Fe] within the cold, dense gas at each average [Fe/H] metallicity. The truncation at higher [Fe/H] abundance is caused by the different time scales for enrichment in the two galaxy patch models. Contours show the fractional mass per bin in this abundance plane and the effects of altering the mass per event while keeping the relative rate ($f_{\rm rp} = 10^{-3}$) constant. Filled in contours represent the simulations with $m_{\rm rp} = \SI{1e-2}{M_\odot}$, while line contours represent $m_{\rm rp} = \SI{1e-1}{M_\odot}$. Results are shown for two galaxy models: the MW progenitor high SFR model (left panel) and satellite galaxy model (right panel). Increasing the mass per event shifts the abundance distributions to higher mean [Eu/Fe] at constant mean [Fe/H], while the abundance spread, which is driven by turbulent mixing, remains unchanged.}
 \label{fig:mass_change}
\end{figure*}

\subsection{The role of $f_{\rm rp}$}
\label{subsec:rate-change}
Our findings in Section~\ref{subsec:mass-change} provide supporting evidence that while the average [Eu/Fe] abundance can be altered by changes in $m_{\rm rp}$, the dispersion of the [Eu/Fe] abundance can only be modified by varying $f_{\rm rp}$. 

It is to this issue that we now turn our attention. Figure~\ref{fig:rate_change} shows how the stacked one-dimensional distributions of [Eu/Fe] as a function of the average [Fe/H] metallicity within the cold, dense gas are altered when $f_{\rm rp}$ is modified. This result should be directly compared with those presented in Figure~\ref{fig:mass_change}. For both of the models considered in Figure~\ref{fig:rate_change}, the relative rate of r-process events is changed from $f_{\rm rp} = 10^{-3}$ to $f_{\rm rp} = 10^{-2}$ while keeping all other inputs constant. 

Some conspicuous points should be underscored from Figure~\ref{fig:rate_change}. First, it is important to note that the [Fe/H] abundance is determined solely by the cc-SNe, whose rate remains unchanged in all simulations. Thus, changes in $f_{\rm rp}$ lead simply to adding more r-process material, thereby increasing the mean [Eu/Fe] abundance but producing very little change in the overall Fe mass. Second, the relative injection rate, $f_{\rm rp}$, alters both the evolution of the mean [Eu/Fe] abundance, as well as, the [Eu/Fe] spread. This is because the total amount of metals injected is significantly increased with time but at the expense of augmenting the number of injection sites and not the total mass per event as was done in in Section~\ref{subsec:mass-change}. As such, the mean separation between injection sites is substantially decreased, which helps the effectiveness of turbulent mixing in smoothing metal inhomogeneities. As expected, this diminishes the [Eu/Fe] abundance spread. Finally, the mean [Eu/Fe] abundance increases more swiftly when $f_{\rm rp}$ is augmented, causing the model to spend more time at higher relative abundances in the plane depicted in Figure~\ref{fig:rate_change}. By its very nature, a mass build up at higher relative abundances is naturally produced in these models. 

Motivated by the results presented in Sections~\ref{subsec:mass-change} and \ref{subsec:rate-change}, we direct our efforts to comparing our models to observations under the assumption that the selected cold gas within the disk is the most likely to form stars.

\begin{figure*}
 \centering
 \includegraphics[width = \textwidth]{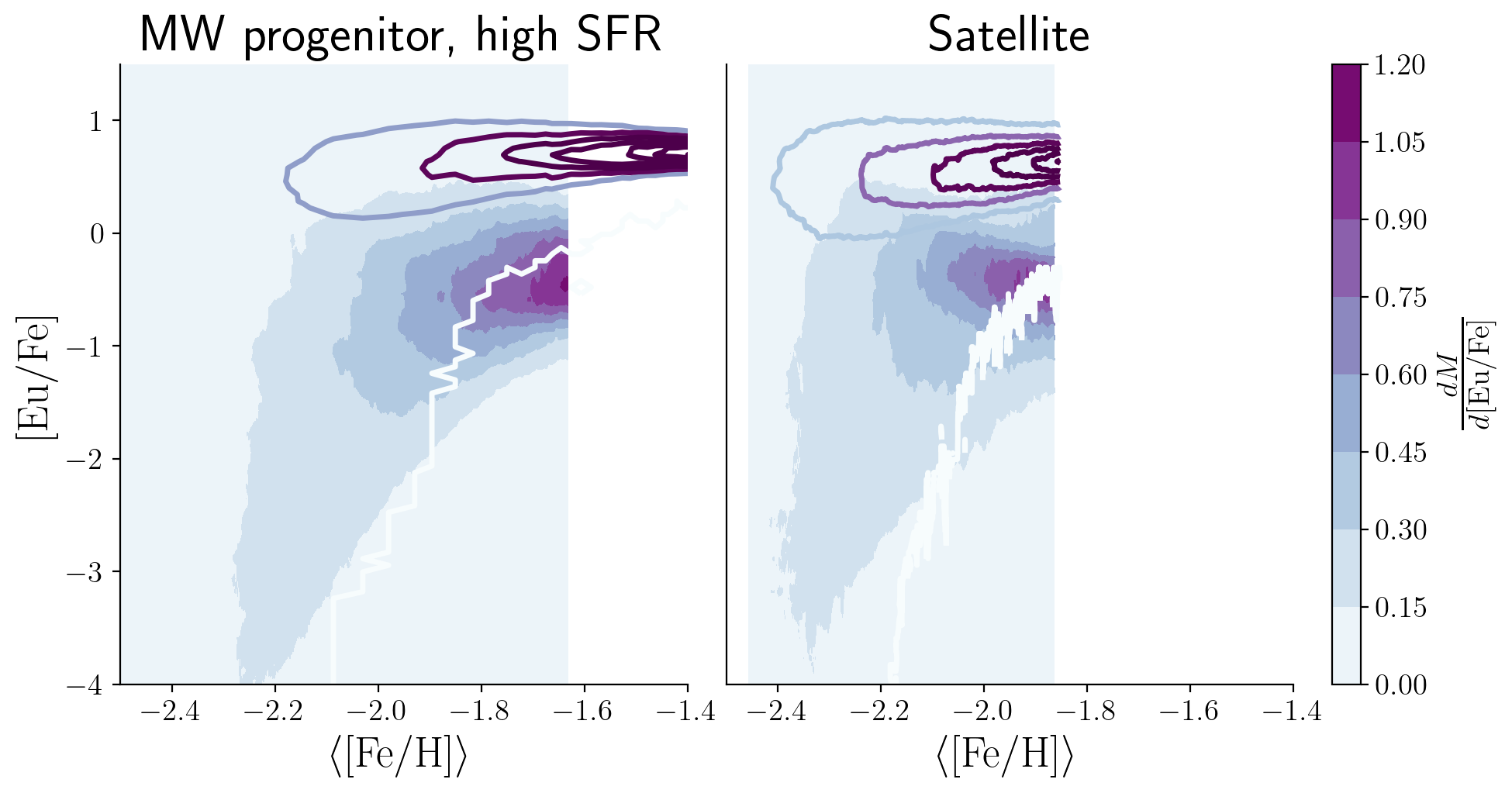}
 \caption{Stacked 1d distributions of [Eu/Fe] within the cold, dense gas at each average [Fe/H] metallicity. The truncation at higher [Fe/H] abundance is caused by the different time scales for enrichment in the two galaxy patches, as well as, slightly different evolution times between different simulations of the same galaxy patch. Results are shown for the same two galaxy models as in Figure~\ref{fig:mass_change}. Color indicates the fractional mass per bin in this abundance plane. Filled in contours represent the simulations with $f_{\rm rp} = 10^{-3}$, while line contours represent $f_{\rm rm} = 10^{-2}$, in both cases $m_{\rm rp } = \SI{1e-2}{M_\odot}$. Changing the relative rate of events from $f_{\rm rp} = 10^{-3}$ to $f_{\rm rp} = 10^{-2}$ reduces the spread of abundances and leads to more cold, dense gas at higher [Eu/Fe] abundances.}
 \label{fig:rate_change}
\end{figure*}

\subsection{Constrains from observations of metal poor halo stars}
\label{subsec:obs_match}

\begin{figure*}
 \centering
 \includegraphics[width = \textwidth]{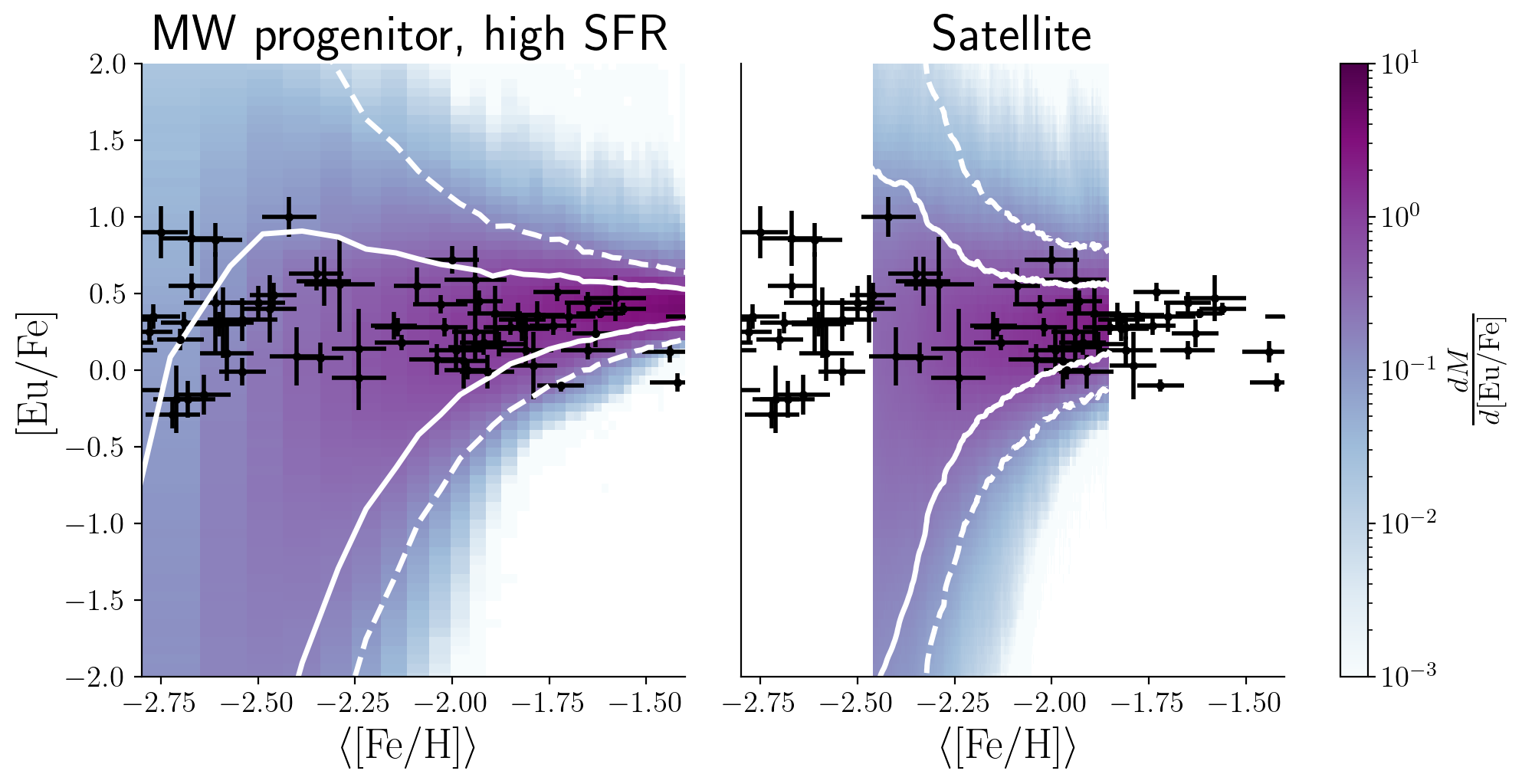}
 \caption{Stacked 1d distributions of [Eu/Fe] in the cold, dense gas at each average [Fe/H] metallicity in the MW progenitor with high SFR model (left panel) and satellite galaxy model (right panel). The color indicates the fractional mass in each bin, the white solid (dashed) lines indicate the $1 \sigma$ ($2 \sigma$) spread of the distribution. In both models, $f_{\rm rp} = 10^{-2}$ and $m_{\rm rp} = \SI{5e-3}\,{M_\odot}$. We consider the cold, dense gas within the disk as the most likely to form stars. The black symbols indicate the abundances of metal poor halo stars with uncertainties as reported by \citet{Roederer2014}. The mean [Eu/Fe] abundance in high SFR model is slightly elevated when compared to the satellite galaxy model as expected from the ability of this model to more effectively retain r-process material. The truncation at higher [Fe/H] abundance is caused by the different time scales for enrichment in the two galaxy patch models. }
 \label{fig:obs_comp}
\end{figure*}
Having considered the influence of $m_{\rm rp}$ and $f_{\rm rp}$ on the mean and spread of [Eu/Fe] as a function of the mean [Fe/H] abundance, we dedicate ourselves to comparing the simulated abundance distributions with the observed stellar abundances in the MW halo. We do this in order to provide useful constraints on the operating conditions of r-process production models in the early Universe. 

The metal-poor halo stars for this analysis are taken from \citet{Roederer2014}. Our selection has a twofold rationale. First, this data set constitutes the largest homogeneously reduced sample of its kind (about 300 metal-poor halo stars). And, second, the authors have systematically and scrupulously documented the uncertainty of their measurements. This allows us to use the star-to-star [Eu/Fe] scatter and mean as a function of [Fe/H] derived by \citet{Roederer2014} as a constraint for r-process production models in the [$m_{\rm rp}$,$f_{\rm rp}$] plane.

In Figure~\ref{fig:obs_comp}, we show the distributions of [Eu/Fe] as a function of the average abundance of [Fe/H] within the cold, dense gas in the disk for both the MW progenitor with high SFR (left panel) and the satellite galaxy model (right panel). The color indicates the fractional mass in a given bin. Also shown are the abundances and measurement uncertainties of the MW halo stars, as reported by \cite{Roederer2014}. The parameters of the r-process events have been adjusted to achieve a reasonable agreement with the observations for both models. 
%To facilitate comparison we use the same values of $m_{\rm rp}$ and $f_{\rm rp}$ and computed a grid of parameters.
Shown is a representative model that effectively recovers both the mean and the spread in [Eu/Fe] as a function of [Fe/H] for the sample of MW halo stars that we selected. The representative values used to generate the models in Figure~\ref{fig:obs_comp} are $f_{\rm rp} = 10^{-2}$ and $m_{\rm rp} = \SI{5e-3}\,{M_\odot}$, respectively. 
Given that the MW progenitor with high SFR model retains the r-process products more effectively, the mean [Eu/Fe] is slightly more elevated compared to the satellite model, albeit within the uncertainties of the data and the range of applicability of the simulations (e.g., results are only shown after steady state is reached). 

From the simple model comparison presented in Figure \ref{fig:obs_comp}, we find clear evidence that the properties of the metal-poor halo stars are consistent with stellar birth sites in a MW progenitor with a high SFR or a satellite galaxy. It is, however, important to note that model comparison is hindered by the fact that the observational sample used in this analysis is impaired by incompleteness and selection effects, which are most evident at low [Eu/Fe] abundances. \\ 
In the preceding sections we outlined how the abundance distributions of [Eu/Fe] are influenced by changes in $m_{\rm rp}$ and $f_{\rm rp}$. In what follows, we briefly consider how these results may be extrapolated beyond the parameter space explored in this paper. The results presented in Section~\ref{subsec:mass-change} indicate that that impact of altering $m_{\rm rp}$ while keeping $f_{\rm rp}$ unaltered is straight forward. Changing $m_{\rm rp}$ only leads to a change in normalization, with an increase of mass per event leading to a larger abundance normalization.\\

Extrapolating $f_{\rm rp}$ is more complicated due the intricate influence of this parameter on the shape of the [Eu/Fe] distribution. The results presented in section~\ref{subsec:rate-change} clearly indicate that lower relative rate leads to later r-process enrichment. At $f_{\rm rp} = 10^{-3}$ the simulations struggle to explain the high levels of r-process pollution of low metallicity stars. For this reason, lower relative rates are unlikely to produce reasonable fits to the observed values. Higher relative rates, i.e. more frequent r-process events, seem more likely to reproduce the observational data at low metallicities. On the other hand, as we have shown here, higher relative rates also lead to more efficient metal mixing, quickly reducing the width of the abundance distribution at higher metallicity values. This makes high relative rates to be unable to reproduce the spread observed in [Eu/Fe]. We expand on these arguments in the following section.

\section{Discussion}
\label{sec:Discussion}
Observations of [Eu/Fe] in metal-poor stars suggest relative yields and variations of yields in r-process production events. Such inherent fluctuations are evident when r-process events inject metals at rates that are significantly reduced when compared to cc-SN rates. These fluctuations are smoothed out by turbulent diffusion driven by cc-SNe, which sets the rate of metal mixing. As we have contended in this paper, the degree of fluctuations along with the mean abundances of [Eu/Fe] are sensitive to the relative mass injection rate of r-process and, to a lesser extent, to the type of galaxy environment. 

\subsection{Production rate of r-process elements}
\label{subsec:Hoto_comp}

\begin{figure*}
 \centering
 \includegraphics[width = \textwidth]{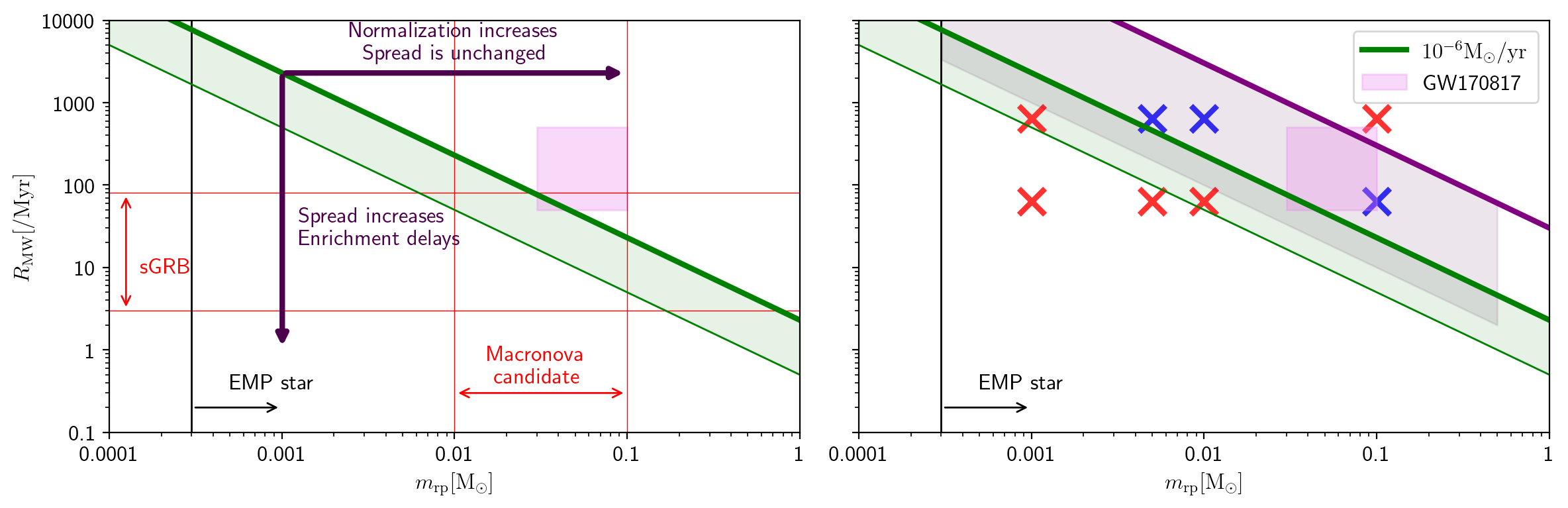}
\caption{The constraints derived for r-process production events in the [$m_{\rm rp}$,~$R_{\rm MW}$] plane, which has been adapted from \citet{Hotokezaka2018}. The left panel includes a compilation of the production rate of r-process element constraints in the MW: the constraints from GW170818, limits on rates from short gamma-ray burst observations, the mass limits from extremely metal-poor stars \citep{2018ApJ...860...89M}, and the mass limits from macronova candidates. The derived relationships between the mean and spread of abundances when the rate or mass per event are altered are highlighted in the left panel. On the right-hand panel we show the constraints derived from turbulent mixing simulations. The red symbols indicate simulations that do not successfully explain the data while the blue symbols are for simulations that give a reasonable description of the [Eu/Fe] abundances (see also Appendix~\ref{app:prod_rate}). The lavender shaded region highlights the parameter space for which simulations provide a reasonable description of the data. The upper limit for the mass production rate (purple line) arise naturally from the highest [Eu/Fe] abundance measurement (see text). Models above this region are stringently ruled out given that there are very few selection effects against uncovering the highest [Eu/Fe] metal poor stars.}
 \label{fig:Hotokezaka_comp}
\end{figure*}

The total r-process mass in the MW is estimated to be $M_{\rm tot, rp} \approx\SI{2.3e4}\,{M_\odot}$ \citep{Hotokezaka2018}. Combining this number with an estimate of the age of the MW ($t_{\rm MW} \approx \SI{1e10}\,{\rm yr}$), one can calculate an average production rate of r-process elements in the galaxy as
\begin{equation}
 \dot{m}_{\rm rp} \approx \frac{M_{\rm tot, rp}}{t_{\rm MW}} \approx \SI{2.3 e-6}{M_\odot \per yr}.
\end{equation}
\citet{Hotokezaka2018} used this relationship to estimate the mass production rate of r-process events in the MW as
\begin{equation}
 R_{\mathrm{MW}} \approx \SI{230}{\per Myr} \left( \frac{m_{\mathrm{rp}}}{\SI{0.01}{M_\odot}} \right)^{-1}.
\end{equation}
This relationship is represented by the thick green line in Figure~\ref{fig:Hotokezaka_comp} along which the production rate of r-process elements is unchanged. The shaded region below the line is limited by the same relationship by for a lower total r-process mass in the MW. This relationship is set by $\dot{m}_{\rm rp} \approx \SI{1e-7}{M_\odot \per yr}$, which is thought to give a strict lower limit to $\dot{m}_{\rm rp}$ based on the minimum amount of r-process mass in stars in the MW \citep{Kasen2017,Hotokezaka2018}. 

The shaded-pink region in Figure~\ref{fig:Hotokezaka_comp} shows the range of values in the [$m_{\rm rp}$, $R_{\rm MW}$] plane derived from observations of GW170817 \citep{Kasen2017,2017Sci...358.1583K,Waxman2018,Hotokezaka2018}. We also show in this plane the estimates of the rates of short gamma-ray bursts ($R_{\rm MW}$), which are believed to be an observational signature of NSM \citep{Eichler1989,2005ApJ...630L.165L,2007NJPh....9...17L} and the mass constraints ($m_{\rm rp}$) from observations of afterglows of sGRBs \citep{2011ApJ...736L..21R,Hotokezaka2013,Tanvir2013,Berger2013,Yang2015,Jin2016,Kasen2015,Kasliwal2017,2017Sci...358.1583K,2017ApJ...848L..34M,2019MNRAS.486..672A}, referred to in Figure~\ref{fig:Hotokezaka_comp} as macronova candidates. 
Finally, the vertical black line marks the lower limit on r-process mass per event as deduced by \citet{2018ApJ...860...89M} based on observations of metal poor halo stars in the MW. 

In the left panel of Figure~\ref{fig:Hotokezaka_comp} we call attention to the results from our turbulent mixing study, which explores how the abundance distribution of [Eu/Fe] is altered by $m_{\rm rp}$ and $f_{\rm rp}$ (i.e., $R_{\rm MW}$). As highlighted by the horizontal arrow, changing $m_{\rm rp}$ while keeping $f_{\rm rp}$ constant shifts the mean abundance at a fixed [Fe/H] abundance but does not alter the scatter. This implies that the observed spread of abundances can not be used to constraint $m_{\rm rp}$. In contrast, as underscored by the vertical arrow, $f_{\rm rp}$ changes both the spread and the mean of the [Eu/Fe] abundance distributions at a fixed [Fe/H] abundance. 

The results of the galaxy patch simulations of the MW progenitor with high SFR model are shown in the right panel of Figure~\ref{fig:Hotokezaka_comp}. To accurately position our simulations in the [$m_{\rm rp}$, ~$R_{\rm MW}$] plane, we follow \citet{Beniamini2016a} who estimates an average rate of cc-SNe in the MW to be 
\begin{equation}
\Tilde{n}_{\rm SNe, MW} \approx \SI{6.4e4}{Myr^{-1}}. 
\end{equation}
This allows us to chart the relative rate of r-process to cc-SN events from the local boxes to standard MW values by assuming constant relative rates across the galaxy. Then it follows that 
\begin{equation}
R_{\rm MW, rp} = f_{\rm rp} \Tilde{n}_{\rm SNe, MW}. 
\end{equation}
From Figures~\ref{fig:mass_change}-\ref{fig:obs_comp} we know that the SFR does not strongly impact the shape of the distribution. As such, we expect that that conclusions based on assuming a constant SFR everywhere in the disk will not significantly be altered if the SFR changes across the disk. \\
We divide the simulations broadly into two distinct groups by how well they describe the observed abundances of MW halo stars \citep{Roederer2014}. We consider both the mean and the spread of [Eu/Fe] abundance in bins of [Fe/H], as well as, the range of [Fe/H] abundance over which there is rough agreement when making these comparisons. In Appendix~\ref{app:prod_rate} we present the results of the simulations with varying $m_{\rm rp}$ and $f_{\rm rp}$ and how they compare to observations. The blue symbols indicate reasonable agreement between observations and simulations, while the red symbols indicate little to no overlap between the abundances predicted by the simulations and those observed in the comparison sample. 

Obviously, the comparison presented in Figure~\ref{fig:Hotokezaka_comp} is only cursory and should be taken as an order of magnitude estimate at present. Having said this, the results of our metal mixing study allow us to clearly define a model confidence region within the [$m_{\rm rp}$, $R_{\rm MW}$] plane, which is specified by the lavender shaded region in the left panel. The lavender region is restrained by a strict upper limit (purple line). Such a limit can be derived by the underlying fact that simulations above this boundary will naturally produce stars with a mean abundance of [Eu/Fe] = 1.4, which is higher than the abundance of the most r-process enhanced star in the \citet{Roederer2014} sample: [Eu/Fe] = 1.37. The slope of the line is derived from the mass ratios of Eu and Fe production leading to [Eu/Fe] = 1.4. This value can be written as a production rate with $\dot{m}_{\rm rp,max} \approx \SI{4.7e-4}{M_\odot \per yr}$. The lavender region presented in the left panel of Figure~\ref{fig:Hotokezaka_comp} is in remarkable good agreement with previous independent constraints derived both theoretically and empirically and provide a significant reduction in the permissible parameter space.

\subsection{Co-production of light and heavy r-process elements?}
\label{subsec:light_heavy}
Several astrophysical processes might be required to explain the solar r-process abundance pattern of both the lighter neutron capture elements (between the first and second r-process peaks at $A\approx 80$ and $A\approx 130$, respectively) and heavier nuclei such as Eu \citep{Cowan2021}. A fascinating characteristic that emerges when comparing the relative abundances of metal poor halo stars is the robustness of the pattern for elements with $A\geqslant 137$ \citep{Sneden2008}. 
In this study we have selected to model the r-process mass contribution as that comprised by all elements with $A \geq 69$. If, for example, we change the minimum atomic number from $A_{\rm min} = 69$ to $A_{\rm min} = 90$ the average abundances of [Eu/Fe] shift by $+0.68$\,dex with no other changes to the results presented in Figure~\ref{fig:Hotokezaka_comp}. Appendix~\ref{app:prod_rate} includes the simulated distributions for $m_{\rm rp}$ calculated with and without this shifting (i.e., with $A_{\rm min} = 69$ and $A_{\rm min} = 90$). This simply implies that constraints on $\dot{m}_{\rm rp}$ are sensitive to whether r-process sites, which are assumed in this paper to be standard, are responsible for producing all r-process elements or only the heavier ones. If Eu is produced by sources that only produce heavier r-process elements ($A \geq 90$), this implies that the derived constraints on $\dot{m}_{\rm rp}$ should be lower by $0.68$\,dex when compared to those illustrated in Figure~\ref{fig:Hotokezaka_comp}. 

\subsection{On the origin of the metal poor halo stars}
\label{subsec:halo-origin}
The MW halo is expected to have been assembled from stars originally residing in disrupted satellites \citep[e.g.][]{Naidu2021,Santistevan2021}. The presence of r-process enhanced stars \citep{Beers2005,Holmbeck2020} in several dwarf galaxy systems and the orbital properties of r-process enhanced stars in the halo\citep{2018AJ....156..179R} indicate that such stars likely were accreted from disrupted satellites. From a theoretical perspective, \citet{Hirai2022} recently performed hydrodynamic zoom-in cosmological simulations, focusing on a MW-like galaxies. 
They found that the vast majority (90\%) of r-process enhanced stars ([Eu/Fe] $> 0.7$) are formed early in the evolution of a galaxy and concluded 
that the majority of these r-process enhanced stars were accreted from disrupted satellite galaxies. 

Our results support the hypothesis that both MW-like and dwarf-like galaxy systems are capable of producing stars with r-process abundances similar to those observed in the MW halo. It is, however, important to note that retention of r-process elements in satellite galaxies is found to be smaller than in MW-like progenitors (Section~\ref{sec:potentials}). This implies that viable mass production rates in satellite galaxies should be larger (by a factor of a few) than those presented in Figure~\ref{fig:Hotokezaka_comp} for a MW-like progenitor. Retention of r-process elements might also be significantly reduced for r-process events with large offsets from the star forming disk, such as expected for NSM with long delay times and large kick velocities. With that said, systems with large displacements are thought to be much less effective at polluting low metallicity gas with r-process material in the early Universe \citep[e.g.,][]{2019Natur.569..241S}.

\section{Summary and conclusions}
\label{sec:conclusions}
In this paper we study the patchy enrichment of the ISM using numerical simulations at kpc scales that are able to resolve the mixing of metals by cc-SNe-driven turbulence. By investigating the statistics of variations of cc-SN and r-process products in these simulations, we are able to derive constraints on the allowed range of the production rate of r-process elements in the MW. By systematically varying the model parameters, we were able to identify some of the physical process that we believe are most relevant to explain the mean and dispersion of [Eu/Fe] abundances in metal-poor stars. Our salient findings are:

\begin{itemize}
\item cc-SNe inject freshly synthesized metals and drive turbulent mixing which causes metals to diffuse across the disk (Figure~\ref{fig:UMW_rhoT_maps}). By virtue of their much more frequent injection, cc-SNe products (for example, Fe) are more evenly distributed than r-process elements (for example, Eu) synthesized in rarer events. This difference naturally produces an ISM with r-process elements that are nonuniform and highly undiluted at early epochs (Figure~\ref{fig:UMW_metal_maps}). These congenital variations are most notable when r-process sources inject metals at rates that are considerably decreased compared to cc-SN rates (Sections \ref{sec:UMW} and \ref{sec:potentials}).

\item The momentum and energy that goes into driving turbulence in the disk also launches galactic winds and flings metals out of the disk (Figure~\ref{fig:UMW_mass_loading}). The rarefied galactic wind is highly enriched in both Fe and Eu compared to the star forming gas in the disk (Figure~\ref{fig:UMW_metal_maps}). This result implies that a considerable mass of r-process elements might reside in the hot inter galactic medium (Sections \ref{sec:UMW} and \ref{sec:potentials}). 

\item The metal mass loading factors of cc-SN and r-process products are not exactly the same, suggesting that r-process metals are launched by winds before they are able to be efficiently mixed with the ISM (Figure~\ref{fig:UMW_metal_maps}). However, models with fewer cc-SNe give r-process elements more time to diffuse and mix in the disk and, as a result, show winds with less prominent Eu-enriched outbursts (Sections \ref{sec:UMW} and \ref{sec:potentials}). 

\item Across different galaxy models, the metal mass loading factors of iron and r-process elements are rather similar. However, the magnitude of the loading factors depend on the specific galaxy potential, with larger mass loading factors found in less massive galaxies (Figure~\ref{fig:galaxy_mass_loading}). Implicitly, higher mass injection rates of r-process material are required in a satellite galaxy in order to achieve an average [Eu/H] abundance comparable with a MW-like progenitor galaxy (Section \ref{sec:potentials}). 

\item The r-process metal mass loading factor shows a measurable increase with increasing allowed height of r-process events (Figure~\ref{fig:rp_sh}). Despite this, the [Eu/Fe] abundances, especially in the cold, dense gas phase, are found to be very similar across models (Figure~\ref{fig:UMW_event_height}). Thus, we conclude that no measurable changes in observed galaxy properties are expected for the range of offsets we are able to probe in these simulations (Section~\ref{sec:zNSM}).

\item The degree of fluctuations in and the mean of [Eu/Fe] abundances in the cold, dense gas are found to be highly responsive to the mass injection rate of r-process (Section~\ref{sec:prod_rate}) and, to a lesser extent, to the type of galaxy (Section~\ref{sec:potentials}). Concretely, increasing the r-process mass per event increases the mean [Eu/Fe] abundance (Figure~\ref{fig:mass_change}), while increasing the rate of r-process events relative to cc-SNe increases the mean [Eu/Fe] abundance and reduces the [Eu/Fe] spread (Figure~\ref{fig:rate_change}). 

\item Observations of [Eu/Fe] in metal-poor stars are used to derive constraints on the mass per event and the event rate of r-process sources (Section~\ref{sec:Discussion}). We find a production rate of $6 \times 10^{-7} {M_\odot / \rm yr} \lesssim \dot{m}_{\rm rp} \ll 4.7 \times 10^{-4} {M_\odot / \rm yr}$ best explains the data. The constraints presented are in notable agreement with other independently derived confinements and produce a marked reduction in the permitted parameter range (Figure~\ref{fig:Hotokezaka_comp}). 

\item Our findings give credence to the idea that stars with r-process abundances similar to those observed in the MW halo can be manufactured by both MW-like and dwarf-like galaxy progenitors (Figure~\ref{fig:obs_comp}). Although we note that
r-process mass retention in satellite galaxies is found to be smaller than in MW-like progenitors and, as such, the viable mass production rates should be appropriately higher (Section~\ref{sec:prod_rate}). 
\end{itemize}

\begin{acknowledgements}
We thank the referee for helpful comments which improved the quality of this paper. We thank C. Sakari, K. Ruiz-Rocha, A. Ji, D. Kasen, N. Imara and L. Kenoly for insightful discussions. This work made use of an HPC facility funded by a grant from VILLUM FONDEN (project number 16599). A.N.K. and E.R.-R. acknowledge support by the Heising-Simons Foundation, the Danish National Research Foundation (DNRF132) and NSF (AST-2206243, AST-1911206 and AST-1852393). PM gratefully acknowledges support from NASA grant 14-WPS14-0048. M.S.F gratefully acknowledges support provided by NASA through Hubble Fellowship grant HST-HF2-51493.001-A awarded by the Space Telescope Science Institute, which is operated by the Association of Universities for Research in Astronomy, In., for NASA, under the contract NAS 5-26555.
\end{acknowledgements}

\bibliographystyle{aasjournal}
\bibliography{references}

\begin{thebibliography}{}
\expandafter\ifx\csname natexlab\endcsname\relax\def\natexlab#1{#1}\fi
\providecommand{\url}[1]{\href{#1}{#1}}
\providecommand{\dodoi}[1]{doi:~\href{http://doi.org/#1}{\nolinkurl{#1}}}
\providecommand{\doeprint}[1]{\href{http://ascl.net/#1}{\nolinkurl{http://ascl.net/#1}}}
\providecommand{\doarXiv}[1]{\href{https://arxiv.org/abs/#1}{\nolinkurl{https://arxiv.org/abs/#1}}}

\bibitem[{{Abbott} {et~al.}(2017){Abbott}, {Abbott}, {Abbott}, {Acernese},
  {Ackley}, {Adams}, {Adams}, {Addesso}, {Adhikari}, {Adya}, {Affeldt},
  {Afrough}, {Agarwal}, {Agathos}, {Agatsuma}, {Aggarwal}, {Aguiar}, {Aiello},
  {Ain}, {Ajith}, {Allen}, {Allen}, {Allocca}, {Altin}, {Amato}, {Ananyeva},
  {Anderson}, {Anderson}, {Angelova}, {Antier}, {Appert}, {Arai}, {Araya},
  {Areeda}, {Arnaud}, {Arun}, {Ascenzi}, {Ashton}, {Ast}, {Aston}, {Astone},
  {Atallah}, {Aufmuth}, {Aulbert}, {AultONeal}, {Austin}, {Avila-Alvarez},
  {Babak}, {Bacon}, {Bader}, {Bae}, {Baker}, {Baldaccini}, {Ballardin},
  {Ballmer}, {Banagiri}, {Barayoga}, {Barclay}, {Barish}, {Barker}, {Barkett},
  {Barone}, {Barr}, {Barsotti}, {Barsuglia}, {Barta}, {Bartlett}, {Bartos},
  {Bassiri}, {Basti}, {Batch}, {Bawaj}, {Bayley}, {Bazzan}, {B{\'e}csy},
  {Beer}, {Bejger}, {Belahcene}, {Bell}, {Berger}, {Bergmann}, {Bero}, {Berry},
  {Bersanetti}, {Bertolini}, {Betzwieser}, {Bhagwat}, {Bhandare}, {Bilenko},
  {Billingsley}, {Billman}, {Birch}, {Birney}, {Birnholtz}, {Biscans},
  {Biscoveanu}, {Bisht}, {Bitossi}, {Biwer}, {Bizouard}, {Blackburn},
  {Blackman}, {Blair}, {Blair}, {Blair}, {Bloemen}, {Bock}, {Bode}, {Boer},
  {Bogaert}, {Bohe}, {Bondu}, {Bonilla}, {Bonnand}, {Boom}, {Bork}, {Boschi},
  {Bose}, {Bossie}, {Bouffanais}, {Bozzi}, {Bradaschia}, {Brady}, {Branchesi},
  {Brau}, {Briant}, {Brillet}, {Brinkmann}, {Brisson}, {Brockill}, {Broida},
  {Brooks}, {Brown}, {Brunett}, {Buchanan}, {Buikema}, {Bulik}, {Bulten},
  {Buonanno}, {Buskulic}, {Buy}, {Byer}, {Cabero}, {Cadonati}, {Cagnoli},
  {Cahillane}, {Calder{\'o}n Bustillo}, {Callister}, {Calloni}, {Camp},
  {Canepa}, {Canizares}, {Cannon}, {Cao}, {Cao}, {Capano}, {Capocasa},
  {Carbognani}, {Caride}, {Carney}, {Casanueva Diaz}, {Casentini}, {Caudill},
  {Cavagli{\`a}}, {Cavalier}, {Cavalieri}, {Cella}, {Cepeda},
  {Cerd{\'a}-Dur{\'a}n}, {Cerretani}, {Cesarini}, {Chamberlin}, {Chan}, {Chao},
  {Charlton}, {Chase}, {Chassande-Mottin}, {Chatterjee}, {Cheeseboro}, {Chen},
  {Chen}, {Chen}, {Cheng}, {Chia}, {Chincarini}, {Chiummo}, {Chmiel}, {Cho},
  {Cho}, {Chow}, {Christensen}, {Chu}, {Chua}, {Chua}, {Chung}, {Chung},
  {Ciani}, {Ciolfi}, {Cirelli}, {Cirone}, {Clara}, {Clark}, {Clearwater},
  {Cleva}, {Cocchieri}, {Coccia}, {Cohadon}, {Cohen}, {Colla}, {Collette},
  {Cominsky}, {Constancio}, {Conti}, {Cooper}, {Corban}, {Corbitt},
  {Cordero-Carri{\'o}n}, {Corley}, {Corsi}, {Cortese}, {Costa}, {Coughlin},
  {Coughlin}, {Coulon}, {Countryman}, {Couvares}, {Covas}, {Cowan}, {Coward},
  {Cowart}, {Coyne}, {Coyne}, {Creighton}, {Creighton}, {Cripe}, {Crowder},
  {Cullen}, {Cumming}, {Cunningham}, {Cuoco}, {Dal Canton}, {D{\'a}lya},
  {Danilishin}, {D'Antonio}, {Danzmann}, {Dasgupta}, {Da Silva Costa},
  {Dattilo}, {Dave}, {Davier}, {Davis}, {Daw}, {Day}, {De}, {DeBra},
  {Degallaix}, {De Laurentis}, {Del{\'e}glise}, {Del Pozzo}, {Demos}, {Denker},
  {Dent}, {De Pietri}, {Dergachev}, {De Rosa}, {DeRosa}, {De Rossi}, {DeSalvo},
  {de Varona}, {Devenson}, {Dhurandhar}, {D{\'\i}az}, {Di Fiore}, {Di
  Giovanni}, {Di Girolamo}, {Di Lieto}, {Di Pace}, {Di Palma}, {Di Renzo},
  {Doctor}, {Dolique}, {Donovan}, {Dooley}, {Doravari}, {Dorrington},
  {Douglas}, {Dovale {\'A}lvarez}, {Downes}, {Drago}, {Dreissigacker},
  {Driggers}, {Du}, {Ducrot}, {Dupej}, {Dwyer}, {Edo}, {Edwards}, {Effler},
  {Eggenstein}, {Ehrens}, {Eichholz}, {Eikenberry}, {Eisenstein}, {Essick},
  {Estevez}, {Etienne}, {Etzel}, {Evans}, {Evans}, {Factourovich}, {Fafone},
  {Fair}, {Fairhurst}, {Fan}, {Farinon}, {Farr}, {Farr}, {Fauchon-Jones},
  {Favata}, {Fays}, {Fee}, {Fehrmann}, {Feicht}, {Fejer}, {Fernandez-Galiana},
  {Ferrante}, {Ferreira}, {Ferrini}, {Fidecaro}, {Finstad}, {Fiori},
  {Fiorucci}, {Fishbach}, {Fisher}, {Fitz-Axen}, {Flaminio}, {Fletcher},
  {Fong}, {Font}, {Forsyth}, {Forsyth}, {Fournier}, {Frasca}, {Frasconi},
  {Frei}, {Freise}, {Frey}, {Frey}, {Fries}, {Fritschel}, {Frolov}, {Fulda},
  {Fyffe}, {Gabbard}, {Gadre}, {Gaebel}, {Gair}, {Gammaitoni}, {Ganija},
  {Gaonkar}, {Garcia-Quiros}, {Garufi}, {Gateley}, {Gaudio}, {Gaur},
  {Gayathri}, {Gehrels}, {Gemme}, {Genin}, {Gennai}, {George}, {George},
  {Gergely}, {Germain}, {Ghonge}, {Ghosh}, {Ghosh}, {Ghosh}, {Giaime},
  {Giardina}, {Giazotto}, {Gill}, {Glover}, {Goetz}, {Goetz}, {Gomes},
  {Goncharov}, {Gonzalez Castro}, {Gopakumar}, {Gorodetsky}, {Gossan},
  {Gosselin}, {Gouaty}, {Grado}, {Graef}, {Granata}, {Grant}, {Gras}, {Gray},
  {Greco}, {Green}, {Gretarsson}, {Groot}, {Grote}, {Grunewald}, {Gruning},
  {Guidi}, {Guo}, {Gupta}, {Gupta}, {Gushwa}, {Gustafson}, {Gustafson},
  {Halim}, {Hall}, {Hall}, {Hamilton}, {Hammond}, {Haney}, {Hanke}, {Hanks},
  {Hanna}, {Hannam}, {Hannuksela}, {Hanson}, {Hardwick}, {Harms}, {Harry},
  {Harry}, {Hart}, {Haster}, {Haughian}, {Healy}, {Heidmann}, {Heintze},
  {Heitmann}, {Hello}, {Hemming}, {Hendry}, {Heng}, {Hennig}, {Heptonstall},
  {Heurs}, {Hild}, {Hinderer}, {Hoak}, {Hofman}, {Holgado}, {Holt}, {Holz},
  {Hopkins}, {Horst}, {Hough}, {Houston}, {Howell}, {Hreibi}, {Hu}, {Huerta},
  {Huet}, {Hughey}, {Husa}, {Huttner}, {Huynh-Dinh}, {Indik}, {Inta}, {Intini},
  {Isa}, {Isac}, {Isi}, {Iyer}, {Izumi}, {Jacqmin}, {Jani}, {Jaranowski},
  {Jawahar}, {Jim{\'e}nez-Forteza}, {Johnson}, {Jones}, {Jones}, {Jonker},
  {Ju}, {Junker}, {Kalaghatgi}, {Kalogera}, {Kamai}, {Kandhasamy}, {Kang},
  {Kanner}, {Kapadia}, {Karki}, {Karvinen}, {Kasprzack}, {Katolik},
  {Katsavounidis}, {Katzman}, {Kaufer}, {Kawabe}, {K{\'e}f{\'e}lian}, {Keitel},
  {Kemball}, {Kennedy}, {Kent}, {Key}, {Khalili}, {Khan}, {Khan}, {Khan},
  {Khazanov}, {Kijbunchoo}, {Kim}, {Kim}, {Kim}, {Kim}, {Kim}, {Kim},
  {Kimball}, {Kimbrell}, {King}, {King}, {Kinley-Hanlon}, {Kirchhoff},
  {Kissel}, {Kleybolte}, {Klimenko}, {Knowles}, {Koch}, {Koehlenbeck}, {Koley},
  {Kondrashov}, {Kontos}, {Korobko}, {Korth}, {Kowalska}, {Kozak},
  {Kr{\"a}mer}, {Kringel}, {Kr{\'o}lak}, {Kuehn}, {Kumar}, {Kumar}, {Kumar},
  {Kuo}, {Kutynia}, {Kwang}, {Lackey}, {Lai}, {Landry}, {Lang}, {Lange},
  {Lantz}, {Lanza}, {Larson}, {Lartaux-Vollard}, {Lasky}, {Laxen}, {Lazzarini},
  {Lazzaro}, {Leaci}, {Leavey}, {Lee}, {Lee}, {Lee}, {Lee}, {Lee}, {Lehmann},
  {Lenon}, {Leonardi}, {Leroy}, {Letendre}, {Levin}, {Li}, {Linker},
  {Littenberg}, {Liu}, {Lo}, {Lockerbie}, {London}, {Lord}, {Lorenzini},
  {Loriette}, {Lormand}, {Losurdo}, {Lough}, {Lousto}, {Lovelace}, {L{\"u}ck},
  {Lumaca}, {Lundgren}, {Lynch}, {Ma}, {Macas}, {Macfoy}, {Machenschalk},
  {MacInnis}, {Macleod}, {Maga{\~n}a Hernandez}, {Maga{\~n}a-Sandoval},
  {Maga{\~n}a Zertuche}, {Magee}, {Majorana}, {Maksimovic}, {Man}, {Mandic},
  {Mangano}, {Mansell}, {Manske}, {Mantovani}, {Marchesoni}, {Marion},
  {M{\'a}rka}, {M{\'a}rka}, {Markakis}, {Markosyan}, {Markowitz}, {Maros},
  {Marquina}, {Martelli}, {Martellini}, {Martin}, {Martin}, {Martynov},
  {Mason}, {Massera}, {Masserot}, {Massinger}, {Masso-Reid}, {Mastrogiovanni},
  {Matas}, {Matichard}, {Matone}, {Mavalvala}, {Mazumder}, {McCarthy},
  {McClelland}, {McCormick}, {McCuller}, {McGuire}, {McIntyre}, {McIver},
  {McManus}, {McNeill}, {McRae}, {McWilliams}, {Meacher}, {Meadors}, {Mehmet},
  {Meidam}, {Mejuto-Villa}, {Melatos}, {Mendell}, {Mercer}, {Merilh},
  {Merzougui}, {Meshkov}, {Messenger}, {Messick}, {Metzdorff}, {Meyers},
  {Miao}, {Michel}, {Middleton}, {Mikhailov}, {Milano}, {Miller}, {Miller},
  {Miller}, {Millhouse}, {Milovich-Goff}, {Minazzoli}, {Minenkov}, {Ming},
  {Mishra}, {Mitra}, {Mitrofanov}, {Mitselmakher}, {Mittleman}, {Moffa},
  {Moggi}, {Mogushi}, {Mohan}, {Mohapatra}, {Montani}, {Moore}, {Moraru},
  {Moreno}, {Morriss}, {Mours}, {Mow-Lowry}, {Mueller}, {Muir}, {Mukherjee},
  {Mukherjee}, {Mukherjee}, {Mukund}, {Mullavey}, {Munch}, {Mu{\~n}iz},
  {Muratore}, {Murray}, {Napier}, {Nardecchia}, {Naticchioni}, {Nayak},
  {Neilson}, {Nelemans}, {Nelson}, {Nery}, {Neunzert}, {Nevin}, {Newport},
  {Newton}, {Ng}, {Nguyen}, {Nichols}, {Nielsen}, {Nissanke}, {Nitz}, {Noack},
  {Nocera}, {Nolting}, {North}, {Nuttall}, {Oberling}, {O'Dea}, {Ogin}, {Oh},
  {Oh}, {Ohme}, {Okada}, {Oliver}, {Oppermann}, {Oram}, {O'Reilly}, {Ormiston},
  {Ortega}, {O'Shaughnessy}, {Ossokine}, {Ottaway}, {Overmier}, {Owen}, {Pace},
  {Page}, {Page}, {Pai}, {Pai}, {Palamos}, {Palashov}, {Palomba}, {Pal-Singh},
  {Pan}, {Pan}, {Pang}, {Pang}, {Pankow}, {Pannarale}, {Pant}, {Paoletti},
  {Paoli}, {Papa}, {Parida}, {Parker}, {Pascucci}, {Pasqualetti},
  {Passaquieti}, {Passuello}, {Patil}, {Patricelli}, {Pearlstone}, {Pedraza},
  {Pedurand}, {Pekowsky}, {Pele}, {Penn}, {Perez}, {Perreca}, {Perri},
  {Pfeiffer}, {Phelps}, {Piccinni}, {Pichot}, {Piergiovanni}, {Pierro},
  {Pillant}, {Pinard}, {Pinto}, {Pirello}, {Pitkin}, {Poe}, {Poggiani},
  {Popolizio}, {Porter}, {Post}, {Powell}, {Prasad}, {Pratt}, {Pratten},
  {Predoi}, {Prestegard}, {Prijatelj}, {Principe}, {Privitera}, {Prodi},
  {Prokhorov}, {Puncken}, {Punturo}, {Puppo}, {P{\"u}rrer}, {Qi}, {Quetschke},
  {Quintero}, {Quitzow-James}, {Rabeling}, {Radkins}, {Raffai}, {Raja},
  {Rajan}, {Rajbhandari}, {Rakhmanov}, {Ramirez}, {Ramos-Buades}, {Rapagnani},
  {Raymond}, {Razzano}, {Read}, {Regimbau}, {Rei}, {Reid}, {Reitze}, {Ren},
  {Reyes}, {Ricci}, {Ricker}, {Rieger}, {Riles}, {Rizzo}, {Robertson}, {Robie},
  {Robinet}, {Rocchi}, {Rolland}, {Rollins}, {Roma}, {Romano}, {Romel},
  {Romie}, {Rosi{\'n}ska}, {Ross}, {Rowan}, {R{\"u}diger}, {Ruggi}, {Rutins},
  {Ryan}, {Sachdev}, {Sadecki}, {Sadeghian}, {Sakellariadou}, {Salconi},
  {Saleem}, {Salemi}, {Samajdar}, {Sammut}, {Sampson}, {Sanchez}, {Sanchez},
  {Sanchis-Gual}, {Sandberg}, {Sanders}, {Sassolas}, {Sathyaprakash}, {Sauter},
  {Savage}, {Sawadsky}, {Schale}, {Scheel}, {Scheuer}, {Schmidt}, {Schmidt},
  {Schnabel}, {Schofield}, {Sch{\"o}nbeck}, {Schreiber}, {Schuette}, {Schulte},
  {Schutz}, {Schwalbe}, {Scott}, {Scott}, {Seidel}, {Sellers}, {Sengupta},
  {Sentenac}, {Sequino}, {Sergeev}, {Shaddock}, {Shaffer}, {Shah}, {Shahriar},
  {Shaner}, {Shao}, {Shapiro}, {Shawhan}, {Sheperd}, {Shoemaker}, {Shoemaker},
  {Siellez}, {Siemens}, {Sieniawska}, {Sigg}, {Silva}, {Singer}, {Singh},
  {Singhal}, {Sintes}, {Slagmolen}, {Smith}, {Smith}, {Smith}, {Somala}, {Son},
  {Sonnenberg}, {Sorazu}, {Sorrentino}, {Souradeep}, {Spencer}, {Srivastava},
  {Staats}, {Staley}, {Steinke}, {Steinlechner}, {Steinlechner}, {Steinmeyer},
  {Stevenson}, {Stone}, {Stops}, {Strain}, {Stratta}, {Strigin}, {Strunk},
  {Sturani}, {Stuver}, {Summerscales}, {Sun}, {Sunil}, {Suresh}, {Sutton},
  {Swinkels}, {Szczepa{\'n}czyk}, {Tacca}, {Tait}, {Talbot}, {Talukder},
  {Tanner}, {T{\'a}pai}, {Taracchini}, {Tasson}, {Taylor}, {Taylor}, {Tewari},
  {Theeg}, {Thies}, {Thomas}, {Thomas}, {Thomas}, {Thorne}, {Thrane}, {Tiwari},
  {Tiwari}, {Tokmakov}, {Toland}, {Tonelli}, {Tornasi}, {Torres-Forn{\'e}},
  {Torrie}, {T{\"o}yr{\"a}}, {Travasso}, {Traylor}, {Trinastic}, {Tringali},
  {Trozzo}, {Tsang}, {Tse}, {Tso}, {Tsukada}, {Tsuna}, {Tuyenbayev}, {Ueno},
  {Ugolini}, {Unnikrishnan}, {Urban}, {Usman}, {Vahlbruch}, {Vajente},
  {Valdes}, {van Bakel}, {van Beuzekom}, {van den Brand}, {Van Den Broeck},
  {Vander-Hyde}, {van der Schaaf}, {van Heijningen}, {van Veggel}, {Vardaro},
  {Varma}, {Vass}, {Vas{\'u}th}, {Vecchio}, {Vedovato}, {Veitch}, {Veitch},
  {Venkateswara}, {Venugopalan}, {Verkindt}, {Vetrano}, {Vicer{\'e}}, {Viets},
  {Vinciguerra}, {Vine}, {Vinet}, {Vitale}, {Vo}, {Vocca}, {Vorvick},
  {Vyatchanin}, {Wade}, {Wade}, {Wade}, {Walet}, {Walker}, {Wallace}, {Walsh},
  {Wang}, {Wang}, {Wang}, {Wang}, {Wang}, {Ward}, {Warner}, {Was}, {Watchi},
  {Weaver}, {Wei}, {Weinert}, {Weinstein}, {Weiss}, {Wen}, {Wessel},
  {We{\ss}els}, {Westerweck}, {Westphal}, {Wette}, {Whelan}, {Whiting},
  {Whittle}, {Wilken}, {Williams}, {Williams}, {Williamson}, {Willis},
  {Willke}, {Wimmer}, {Winkler}, {Wipf}, {Wittel}, {Woan}, {Woehler},
  {Wofford}, {Wong}, {Worden}, {Wright}, {Wu}, {Wysocki}, {Xiao}, {Yamamoto},
  {Yancey}, {Yang}, {Yap}, {Yazback}, {Yu}, {Yu}, {Yvert}, {Zadro{\.z}ny},
  {Zanolin}, {Zelenova}, {Zendri}, {Zevin}, {Zhang}, {Zhang}, {Zhang}, {Zhang},
  {Zhao}, {Zhou}, {Zhou}, {Zhu}, {Zhu}, {Zucker}, {Zweizig}, {(LIGO Scientific
  Collaboration}, \& {Virgo Collaboration}}]{Abbott2017}
{Abbott}, B.~P., {Abbott}, R., {Abbott}, T.~D., {et~al.} 2017, \apjl, 850, L40,
  \dodoi{10.3847/2041-8213/aa93fc}

\bibitem[{{Ascenzi} {et~al.}(2019){Ascenzi}, {Coughlin}, {Dietrich}, {Foley},
  {Ramirez-Ruiz}, {Piranomonte}, {Mockler}, {Murguia-Berthier}, {Fryer},
  {Lloyd-Ronning}, \& {Rosswog}}]{2019MNRAS.486..672A}
{Ascenzi}, S., {Coughlin}, M.~W., {Dietrich}, T., {et~al.} 2019, \mnras, 486,
  672, \dodoi{10.1093/mnras/stz891}

\bibitem[{Asplund {et~al.}(2009)Asplund, Grevesse, Sauval, \&
  Scott}]{Asplund2009}
Asplund, M., Grevesse, N., Sauval, A.~J., \& Scott, P. 2009, Annu. Rev. Astron.
  Astrophys., 47, 481, \dodoi{10.1146/annurev.astro.46.060407.145222}

\bibitem[{{Beers} \& {Christlieb}(2005)}]{Beers2005}
{Beers}, T.~C., \& {Christlieb}, N. 2005, \araa, 43, 531,
  \dodoi{10.1146/annurev.astro.42.053102.134057}

\bibitem[{{Behroozi} {et~al.}(2014){Behroozi}, {Ramirez-Ruiz}, \&
  {Fryer}}]{2014ApJ...792..123B}
{Behroozi}, P.~S., {Ramirez-Ruiz}, E., \& {Fryer}, C.~L. 2014, \apj, 792, 123,
  \dodoi{10.1088/0004-637X/792/2/123}

\bibitem[{{Beniamini} {et~al.}(2016){Beniamini}, {Hotokezaka}, \&
  {Piran}}]{Beniamini2016a}
{Beniamini}, P., {Hotokezaka}, K., \& {Piran}, T. 2016, \apj, 832, 149,
  \dodoi{10.3847/0004-637X/832/2/149}

\bibitem[{{Berger} {et~al.}(2013){Berger}, {Fong}, \& {Chornock}}]{Berger2013}
{Berger}, E., {Fong}, W., \& {Chornock}, R. 2013, \apjl, 774, L23,
  \dodoi{10.1088/2041-8205/774/2/L23}

\bibitem[{{Burbidge} {et~al.}(1957){Burbidge}, {Burbidge}, {Fowler}, \&
  {Hoyle}}]{1957RvMP...29..547B}
{Burbidge}, E.~M., {Burbidge}, G.~R., {Fowler}, W.~A., \& {Hoyle}, F. 1957,
  Reviews of Modern Physics, 29, 547, \dodoi{10.1103/RevModPhys.29.547}

\bibitem[{{Cameron}(1957)}]{1957PASP...69..201C}
{Cameron}, A.~G.~W. 1957, \pasp, 69, 201, \dodoi{10.1086/127051}

\bibitem[{{Cioffi} {et~al.}(1988){Cioffi}, {McKee}, \&
  {Bertschinger}}]{Cioffi1988}
{Cioffi}, D.~F., {McKee}, C.~F., \& {Bertschinger}, E. 1988, \apj, 334, 252,
  \dodoi{10.1086/166834}

\bibitem[{{C{\^o}t{\'e}} {et~al.}(2019){C{\^o}t{\'e}}, {Eichler}, {Arcones},
  {Hansen}, {Simonetti}, {Frebel}, {Fryer}, {Pignatari}, {Reichert},
  {Belczynski}, \& {Matteucci}}]{2019ApJ...875..106C}
{C{\^o}t{\'e}}, B., {Eichler}, M., {Arcones}, A., {et~al.} 2019, \apj, 875,
  106, \dodoi{10.3847/1538-4357/ab10db}

\bibitem[{{Coulter} {et~al.}(2017){Coulter}, {Foley}, {Kilpatrick}, {Drout},
  {Piro}, {Shappee}, {Siebert}, {Simon}, {Ulloa}, {Kasen}, {Madore},
  {Murguia-Berthier}, {Pan}, {Prochaska}, {Ramirez-Ruiz}, {Rest}, \&
  {Rojas-Bravo}}]{2017Sci...358.1556C}
{Coulter}, D.~A., {Foley}, R.~J., {Kilpatrick}, C.~D., {et~al.} 2017, Science,
  358, 1556, \dodoi{10.1126/science.aap9811}

\bibitem[{{Cowan} \& {Sneden}(2006)}]{Cowan2006}
{Cowan}, J.~J., \& {Sneden}, C. 2006, \nat, 440, 1151,
  \dodoi{10.1038/nature04807}

\bibitem[{{Cowan} {et~al.}(2021){Cowan}, {Sneden}, {Lawler}, {Aprahamian},
  {Wiescher}, {Langanke}, {Mart{\'\i}nez-Pinedo}, \& {Thielemann}}]{Cowan2021}
{Cowan}, J.~J., {Sneden}, C., {Lawler}, J.~E., {et~al.} 2021, Reviews of Modern
  Physics, 93, 015002, \dodoi{10.1103/RevModPhys.93.015002}

\bibitem[{{Cowan} \& {Thielemann}(2004)}]{2004PhT....57j..47C}
{Cowan}, J.~J., \& {Thielemann}, F.-K. 2004, Physics Today, 57, 10.47,
  \dodoi{10.1063/1.1825268}

\bibitem[{{Eichler} {et~al.}(1989){Eichler}, {Livio}, {Piran}, \&
  {Schramm}}]{Eichler1989}
{Eichler}, D., {Livio}, M., {Piran}, T., \& {Schramm}, D.~N. 1989, \nat, 340,
  126, \dodoi{10.1038/340126a0}

\bibitem[{{Fielding} {et~al.}(2017){Fielding}, {Quataert}, {Martizzi}, \&
  {Faucher-Gigu{\`e}re}}]{2017MNRAS.470L..39F}
{Fielding}, D., {Quataert}, E., {Martizzi}, D., \& {Faucher-Gigu{\`e}re}, C.-A.
  2017, \mnras, 470, L39, \dodoi{10.1093/mnrasl/slx072}

\bibitem[{{Fields} {et~al.}(2002){Fields}, {Truran}, \&
  {Cowan}}]{2002ApJ...575..845F}
{Fields}, B.~D., {Truran}, J.~W., \& {Cowan}, J.~J. 2002, \apj, 575, 845,
  \dodoi{10.1086/341331}

\bibitem[{{Haynes} \& {Kobayashi}(2019)}]{Haynes2019}
{Haynes}, C.~J., \& {Kobayashi}, C. 2019, \mnras, 483, 5123,
  \dodoi{10.1093/mnras/sty3389}

\bibitem[{{Hirai} {et~al.}(2022){Hirai}, {Beers}, {Chiba}, {Aoki}, {Shank},
  {Saitoh}, {Okamoto}, \& {Makino}}]{Hirai2022}
{Hirai}, Y., {Beers}, T.~C., {Chiba}, M., {et~al.} 2022, \mnras, 517, 4856,
  \dodoi{10.1093/mnras/stac2489}

\bibitem[{{Holmbeck} {et~al.}(2020){Holmbeck}, {Hansen}, {Beers}, {Placco},
  {Whitten}, {Rasmussen}, {Roederer}, {Ezzeddine}, {Sakari}, {Frebel}, {Drout},
  {Simon}, {Thompson}, {Bland-Hawthorn}, {Gibson}, {Grebel}, {Kordopatis},
  {Kunder}, {Mel{\'e}ndez}, {Navarro}, {Reid}, {Seabroke}, {Steinmetz},
  {Watson}, \& {Wyse}}]{Holmbeck2020}
{Holmbeck}, E.~M., {Hansen}, T.~T., {Beers}, T.~C., {et~al.} 2020, \apjs, 249,
  30, \dodoi{10.3847/1538-4365/ab9c19}

\bibitem[{{Hotokezaka} {et~al.}(2018){Hotokezaka}, {Beniamini}, \&
  {Piran}}]{Hotokezaka2018}
{Hotokezaka}, K., {Beniamini}, P., \& {Piran}, T. 2018, International Journal
  of Modern Physics D, 27, 1842005, \dodoi{10.1142/S0218271818420051}

\bibitem[{{Hotokezaka} {et~al.}(2013){Hotokezaka}, {Kyutoku}, {Tanaka},
  {Kiuchi}, {Sekiguchi}, {Shibata}, \& {Wanajo}}]{Hotokezaka2013}
{Hotokezaka}, K., {Kyutoku}, K., {Tanaka}, M., {et~al.} 2013, \apjl, 778, L16,
  \dodoi{10.1088/2041-8205/778/1/L16}

\bibitem[{{Ji} {et~al.}(2016{\natexlab{a}}){Ji}, {Frebel}, {Chiti}, \&
  {Simon}}]{Ji2016a}
{Ji}, A.~P., {Frebel}, A., {Chiti}, A., \& {Simon}, J.~D. 2016{\natexlab{a}},
  \nat, 531, 610, \dodoi{10.1038/nature17425}

\bibitem[{{Ji} {et~al.}(2016{\natexlab{b}}){Ji}, {Frebel}, {Simon}, \&
  {Chiti}}]{Ji2016b}
{Ji}, A.~P., {Frebel}, A., {Simon}, J.~D., \& {Chiti}, A. 2016{\natexlab{b}},
  \apj, 830, 93, \dodoi{10.3847/0004-637X/830/2/93}

\bibitem[{{Ji} {et~al.}(2022){Ji}, {Simon}, {Roederer}, {Magg}, {Frebel},
  {Johnson}, {Klessen}, {Magg}, {Cescutti}, {Mateo}, {Bergemann}, \&
  {Bailey}}]{Ji2022}
{Ji}, A.~P., {Simon}, J.~D., {Roederer}, I.~U., {et~al.} 2022, arXiv e-prints,
  arXiv:2207.03499.
\newblock \doarXiv{2207.03499}

\bibitem[{{Jin} {et~al.}(2016){Jin}, {Hotokezaka}, {Li}, {Tanaka}, {D'Avanzo},
  {Fan}, {Covino}, {Wei}, \& {Piran}}]{Jin2016}
{Jin}, Z.-P., {Hotokezaka}, K., {Li}, X., {et~al.} 2016, Nature Communications,
  7, 12898, \dodoi{10.1038/ncomms12898}

\bibitem[{{Karpov} {et~al.}(2020){Karpov}, {Martizzi}, {Macias},
  {Ramirez-Ruiz}, {Kolborg}, \& {Naiman}}]{2020ApJ...896...66K}
{Karpov}, P.~I., {Martizzi}, D., {Macias}, P., {et~al.} 2020, \apj, 896, 66,
  \dodoi{10.3847/1538-4357/ab8f23}

\bibitem[{{Kasen} {et~al.}(2015){Kasen}, {Fern{\'a}ndez}, \&
  {Metzger}}]{Kasen2015}
{Kasen}, D., {Fern{\'a}ndez}, R., \& {Metzger}, B.~D. 2015, \mnras, 450, 1777,
  \dodoi{10.1093/mnras/stv721}

\bibitem[{{Kasen} {et~al.}(2017){Kasen}, {Metzger}, {Barnes}, {Quataert}, \&
  {Ramirez-Ruiz}}]{Kasen2017}
{Kasen}, D., {Metzger}, B., {Barnes}, J., {Quataert}, E., \& {Ramirez-Ruiz}, E.
  2017, \nat, 551, 80, \dodoi{10.1038/nature24453}

\bibitem[{{Kasliwal} {et~al.}(2017){Kasliwal}, {Korobkin}, {Lau}, {Wollaeger},
  \& {Fryer}}]{Kasliwal2017}
{Kasliwal}, M.~M., {Korobkin}, O., {Lau}, R.~M., {Wollaeger}, R., \& {Fryer},
  C.~L. 2017, \apjl, 843, L34, \dodoi{10.3847/2041-8213/aa799d}

\bibitem[{{Kelley} {et~al.}(2010){Kelley}, {Ramirez-Ruiz}, {Zemp}, {Diemand},
  \& {Mandel}}]{2010ApJ...725L..91K}
{Kelley}, L.~Z., {Ramirez-Ruiz}, E., {Zemp}, M., {Diemand}, J., \& {Mandel}, I.
  2010, \apjl, 725, L91, \dodoi{10.1088/2041-8205/725/1/L91}

\bibitem[{{Kilpatrick} {et~al.}(2017){Kilpatrick}, {Foley}, {Kasen},
  {Murguia-Berthier}, {Ramirez-Ruiz}, {Coulter}, {Drout}, {Piro}, {Shappee},
  {Boutsia}, {Contreras}, {Di Mille}, {Madore}, {Morrell}, {Pan}, {Prochaska},
  {Rest}, {Rojas-Bravo}, {Siebert}, {Simon}, \& {Ulloa}}]{2017Sci...358.1583K}
{Kilpatrick}, C.~D., {Foley}, R.~J., {Kasen}, D., {et~al.} 2017, Science, 358,
  1583, \dodoi{10.1126/science.aaq0073}

\bibitem[{{Kolborg} {et~al.}(2022){Kolborg}, {Martizzi}, {Ramirez-Ruiz},
  {Pfister}, {Sakari}, {Wechsler}, \& {Soares-Furtado}}]{Kolborg2022}
{Kolborg}, A.~N., {Martizzi}, D., {Ramirez-Ruiz}, E., {et~al.} 2022, \apjl,
  936, L26, \dodoi{10.3847/2041-8213/ac8c98}

\bibitem[{{Kuijken} \& {Gilmore}(1989)}]{Kuijken1989}
{Kuijken}, K., \& {Gilmore}, G. 1989, \mnras, 239, 605,
  \dodoi{10.1093/mnras/239.2.605}

\bibitem[{{Lee} \& {Ramirez-Ruiz}(2007)}]{2007NJPh....9...17L}
{Lee}, W.~H., \& {Ramirez-Ruiz}, E. 2007, New Journal of Physics, 9, 17,
  \dodoi{10.1088/1367-2630/9/1/017}

\bibitem[{{Lee} {et~al.}(2005){Lee}, {Ramirez-Ruiz}, \&
  {Granot}}]{2005ApJ...630L.165L}
{Lee}, W.~H., {Ramirez-Ruiz}, E., \& {Granot}, J. 2005, \apjl, 630, L165,
  \dodoi{10.1086/496882}

\bibitem[{{Li} \& {Bryan}(2020)}]{Li2020}
{Li}, M., \& {Bryan}, G.~L. 2020, \apjl, 890, L30,
  \dodoi{10.3847/2041-8213/ab7304}

\bibitem[{{Macias} \& {Ramirez-Ruiz}(2018)}]{2018ApJ...860...89M}
{Macias}, P., \& {Ramirez-Ruiz}, E. 2018, \apj, 860, 89,
  \dodoi{10.3847/1538-4357/aac3e0}

\bibitem[{{Macias} \& {Ramirez-Ruiz}(2019)}]{Macias2019}
---. 2019, \apjl, 877, L24, \dodoi{10.3847/2041-8213/ab2049}

\bibitem[{{Martizzi} {et~al.}(2015){Martizzi}, {Faucher-Gigu{\`e}re}, \&
  {Quataert}}]{Martizzi2015}
{Martizzi}, D., {Faucher-Gigu{\`e}re}, C.-A., \& {Quataert}, E. 2015, \mnras,
  450, 504, \dodoi{10.1093/mnras/stv562}

\bibitem[{{Martizzi} {et~al.}(2016){Martizzi}, {Fielding},
  {Faucher-Gigu{\`e}re}, \& {Quataert}}]{Martizzi2016}
{Martizzi}, D., {Fielding}, D., {Faucher-Gigu{\`e}re}, C.-A., \& {Quataert}, E.
  2016, \mnras, 459, 2311, \dodoi{10.1093/mnras/stw745}

\bibitem[{{McWilliam} {et~al.}(1995){McWilliam}, {Preston}, {Sneden}, \&
  {Searle}}]{1995AJ....109.2757M}
{McWilliam}, A., {Preston}, G.~W., {Sneden}, C., \& {Searle}, L. 1995, \aj,
  109, 2757, \dodoi{10.1086/117486}

\bibitem[{{Metzger} {et~al.}(2010){Metzger}, {Mart{\'\i}nez-Pinedo}, {Darbha},
  {Quataert}, {Arcones}, {Kasen}, {Thomas}, {Nugent}, {Panov}, \&
  {Zinner}}]{2010MNRAS.406.2650M}
{Metzger}, B.~D., {Mart{\'\i}nez-Pinedo}, G., {Darbha}, S., {et~al.} 2010,
  \mnras, 406, 2650, \dodoi{10.1111/j.1365-2966.2010.16864.x}

\bibitem[{{Montes} {et~al.}(2016){Montes}, {Ramirez-Ruiz}, {Naiman}, {Shen}, \&
  {Lee}}]{2016ApJ...830...12M}
{Montes}, G., {Ramirez-Ruiz}, E., {Naiman}, J., {Shen}, S., \& {Lee}, W.~H.
  2016, \apj, 830, 12, \dodoi{10.3847/0004-637X/830/1/12}

\bibitem[{{M{\"o}sta} {et~al.}(2018){M{\"o}sta}, {Roberts}, {Halevi}, {Ott},
  {Lippuner}, {Haas}, \& {Schnetter}}]{2018ApJ...864..171M}
{M{\"o}sta}, P., {Roberts}, L.~F., {Halevi}, G., {et~al.} 2018, \apj, 864, 171,
  \dodoi{10.3847/1538-4357/aad6ec}

\bibitem[{{Murguia-Berthier} {et~al.}(2017){Murguia-Berthier}, {Ramirez-Ruiz},
  {Kilpatrick}, {Foley}, {Kasen}, {Lee}, {Piro}, {Coulter}, {Drout}, {Madore},
  {Shappee}, {Pan}, {Prochaska}, {Rest}, {Rojas-Bravo}, {Siebert}, \&
  {Simon}}]{2017ApJ...848L..34M}
{Murguia-Berthier}, A., {Ramirez-Ruiz}, E., {Kilpatrick}, C.~D., {et~al.} 2017,
  \apjl, 848, L34, \dodoi{10.3847/2041-8213/aa91b3}

\bibitem[{{Naidu} {et~al.}(2021){Naidu}, {Conroy}, {Bonaca}, {Zaritsky},
  {Weinberger}, {Ting}, {Caldwell}, {Tacchella}, {Han}, {Speagle}, \&
  {Cargile}}]{Naidu2021}
{Naidu}, R.~P., {Conroy}, C., {Bonaca}, A., {et~al.} 2021, \apj, 923, 92,
  \dodoi{10.3847/1538-4357/ac2d2d}

\bibitem[{{Naiman} {et~al.}(2018){Naiman}, {Pillepich}, {Springel},
  {Ramirez-Ruiz}, {Torrey}, {Vogelsberger}, {Pakmor}, {Nelson}, {Marinacci},
  {Hernquist}, {Weinberger}, \& {Genel}}]{Naiman2018}
{Naiman}, J.~P., {Pillepich}, A., {Springel}, V., {et~al.} 2018, \mnras, 477,
  1206, \dodoi{10.1093/mnras/sty618}

\bibitem[{{Nishimura} {et~al.}(2015){Nishimura}, {Takiwaki}, \&
  {Thielemann}}]{2015ApJ...810..109N}
{Nishimura}, N., {Takiwaki}, T., \& {Thielemann}, F.-K. 2015, \apj, 810, 109,
  \dodoi{10.1088/0004-637X/810/2/109}

\bibitem[{{Ramirez-Ruiz} \& {MacFadyen}(2010)}]{2010ApJ...716.1028R}
{Ramirez-Ruiz}, E., \& {MacFadyen}, A.~I. 2010, \apj, 716, 1028,
  \dodoi{10.1088/0004-637X/716/2/1028}

\bibitem[{{Ramirez-Ruiz} {et~al.}(2015){Ramirez-Ruiz}, {Trenti}, {MacLeod},
  {Roberts}, {Lee}, \& {Saladino-Rosas}}]{2015ApJ...802L..22R}
{Ramirez-Ruiz}, E., {Trenti}, M., {MacLeod}, M., {et~al.} 2015, \apjl, 802,
  L22, \dodoi{10.1088/2041-8205/802/2/L22}

\bibitem[{{Roberts} {et~al.}(2011){Roberts}, {Kasen}, {Lee}, \&
  {Ramirez-Ruiz}}]{2011ApJ...736L..21R}
{Roberts}, L.~F., {Kasen}, D., {Lee}, W.~H., \& {Ramirez-Ruiz}, E. 2011, \apjl,
  736, L21, \dodoi{10.1088/2041-8205/736/1/L21}

\bibitem[{{Roederer} {et~al.}(2018){Roederer}, {Hattori}, \&
  {Valluri}}]{2018AJ....156..179R}
{Roederer}, I.~U., {Hattori}, K., \& {Valluri}, M. 2018, \aj, 156, 179,
  \dodoi{10.3847/1538-3881/aadd9c}

\bibitem[{{Roederer} {et~al.}(2014){Roederer}, {Preston}, {Thompson},
  {Shectman}, \& {Sneden}}]{Roederer2014}
{Roederer}, I.~U., {Preston}, G.~W., {Thompson}, I.~B., {Shectman}, S.~A., \&
  {Sneden}, C. 2014, \apj, 784, 158, \dodoi{10.1088/0004-637X/784/2/158}

\bibitem[{{Roederer} {et~al.}(2016){Roederer}, {Mateo}, {Bailey}, {Song},
  {Bell}, {Crane}, {Loebman}, {Nidever}, {Olszewski}, {Shectman}, {Thompson},
  {Valluri}, \& {Walker}}]{Roederer2016}
{Roederer}, I.~U., {Mateo}, M., {Bailey}, John~I., I., {et~al.} 2016, \aj, 151,
  82, \dodoi{10.3847/0004-6256/151/3/82}

\bibitem[{{Rosswog} \& {Ramirez-Ruiz}(2002)}]{2002MNRAS.336L...7R}
{Rosswog}, S., \& {Ramirez-Ruiz}, E. 2002, \mnras, 336, L7,
  \dodoi{10.1046/j.1365-8711.2002.05898.x}

\bibitem[{{Rosswog} {et~al.}(2003){Rosswog}, {Ramirez-Ruiz}, \&
  {Davies}}]{2003MNRAS.345.1077R}
{Rosswog}, S., {Ramirez-Ruiz}, E., \& {Davies}, M.~B. 2003, \mnras, 345, 1077,
  \dodoi{10.1046/j.1365-2966.2003.07032.x}

\bibitem[{{Safarzadeh} {et~al.}(2019){Safarzadeh}, {Ramirez-Ruiz}, {Andrews},
  {Macias}, {Fragos}, \& {Scannapieco}}]{2019ApJ...872..105S}
{Safarzadeh}, M., {Ramirez-Ruiz}, E., {Andrews}, J.~J., {et~al.} 2019, \apj,
  872, 105, \dodoi{10.3847/1538-4357/aafe0e}

\bibitem[{{Santistevan} {et~al.}(2021){Santistevan}, {Wetzel}, {Sanderson},
  {El-Badry}, {Samuel}, \& {Faucher-Gigu{\`e}re}}]{Santistevan2021}
{Santistevan}, I.~B., {Wetzel}, A., {Sanderson}, R.~E., {et~al.} 2021, \mnras,
  505, 921, \dodoi{10.1093/mnras/stab1345}

\bibitem[{{Shen} {et~al.}(2015){Shen}, {Cooke}, {Ramirez-Ruiz}, {Madau},
  {Mayer}, \& {Guedes}}]{Shen2015}
{Shen}, S., {Cooke}, R.~J., {Ramirez-Ruiz}, E., {et~al.} 2015, \apj, 807, 115,
  \dodoi{10.1088/0004-637X/807/2/115}

\bibitem[{{Siegel} {et~al.}(2019){Siegel}, {Barnes}, \&
  {Metzger}}]{2019Natur.569..241S}
{Siegel}, D.~M., {Barnes}, J., \& {Metzger}, B.~D. 2019, \nat, 569, 241,
  \dodoi{10.1038/s41586-019-1136-0}

\bibitem[{{Sneden} {et~al.}(2008){Sneden}, {Cowan}, \& {Gallino}}]{Sneden2008}
{Sneden}, C., {Cowan}, J.~J., \& {Gallino}, R. 2008, \araa, 46, 241,
  \dodoi{10.1146/annurev.astro.46.060407.145207}

\bibitem[{{Sneden} {et~al.}(2003){Sneden}, {Cowan}, {Lawler}, {Ivans},
  {Burles}, {Beers}, {Primas}, {Hill}, {Truran}, {Fuller}, {Pfeiffer}, \&
  {Kratz}}]{Sneden2003}
{Sneden}, C., {Cowan}, J.~J., {Lawler}, J.~E., {et~al.} 2003, \apj, 591, 936,
  \dodoi{10.1086/375491}

\bibitem[{{Tanvir} {et~al.}(2013){Tanvir}, {Levan}, {Fruchter}, {Hjorth},
  {Hounsell}, {Wiersema}, \& {Tunnicliffe}}]{Tanvir2013}
{Tanvir}, N.~R., {Levan}, A.~J., {Fruchter}, A.~S., {et~al.} 2013, \nat, 500,
  547, \dodoi{10.1038/nature12505}

\bibitem[{{Teyssier}(2002)}]{Teyssier2002}
{Teyssier}, R. 2002, \aap, 385, 337, \dodoi{10.1051/0004-6361:20011817}

\bibitem[{{Thielemann} {et~al.}(2017){Thielemann}, {Eichler}, {Panov}, \&
  {Wehmeyer}}]{Thielmann2017}
{Thielemann}, F.~K., {Eichler}, M., {Panov}, I.~V., \& {Wehmeyer}, B. 2017,
  Annual Review of Nuclear and Particle Science, 67, 253,
  \dodoi{10.1146/annurev-nucl-101916-123246}

\bibitem[{{Thornton} {et~al.}(1998){Thornton}, {Gaudlitz}, {Janka}, \&
  {Steinmetz}}]{Thornton1998}
{Thornton}, K., {Gaudlitz}, M., {Janka}, H.~T., \& {Steinmetz}, M. 1998, \apj,
  500, 95, \dodoi{10.1086/305704}

\bibitem[{{van de Voort} {et~al.}(2022){van de Voort}, {Pakmor}, {Bieri}, \&
  {Grand}}]{VandeVoort2022}
{van de Voort}, F., {Pakmor}, R., {Bieri}, R., \& {Grand}, R. J.~J. 2022,
  \mnras, 512, 5258, \dodoi{10.1093/mnras/stac710}

\bibitem[{{van de Voort} {et~al.}(2020){van de Voort}, {Pakmor}, {Grand},
  {Springel}, {G{\'o}mez}, \& {Marinacci}}]{VandeVoort2020}
{van de Voort}, F., {Pakmor}, R., {Grand}, R. J.~J., {et~al.} 2020, \mnras,
  494, 4867, \dodoi{10.1093/mnras/staa754}

\bibitem[{{van de Voort} {et~al.}(2015){van de Voort}, {Quataert}, {Hopkins},
  {Kere{\v{s}}}, \& {Faucher-Gigu{\`e}re}}]{VandeVoort2015}
{van de Voort}, F., {Quataert}, E., {Hopkins}, P.~F., {Kere{\v{s}}}, D., \&
  {Faucher-Gigu{\`e}re}, C.-A. 2015, \mnras, 447, 140,
  \dodoi{10.1093/mnras/stu2404}

\bibitem[{{Wanajo} {et~al.}(2021){Wanajo}, {Hirai}, \& {Prantzos}}]{Wanajo2021}
{Wanajo}, S., {Hirai}, Y., \& {Prantzos}, N. 2021, \mnras, 505, 5862,
  \dodoi{10.1093/mnras/stab1655}

\bibitem[{{Wang} {et~al.}(2021){Wang}, {Nadler}, {Mao}, {Adhikari}, {Wechsler},
  \& {Behroozi}}]{2021ApJ...915..116W}
{Wang}, Y., {Nadler}, E.~O., {Mao}, Y.-Y., {et~al.} 2021, \apj, 915, 116,
  \dodoi{10.3847/1538-4357/ac024a}

\bibitem[{{Wasserburg} \& {Qian}(2000)}]{2000ApJ...529L..21W}
{Wasserburg}, G.~J., \& {Qian}, Y.~Z. 2000, \apjl, 529, L21,
  \dodoi{10.1086/312455}

\bibitem[{{Watson} {et~al.}(2019){Watson}, {Hansen}, {Selsing}, {Koch},
  {Malesani}, {Andersen}, {Fynbo}, {Arcones}, {Bauswein}, {Covino}, {Grado},
  {Heintz}, {Hunt}, {Kouveliotou}, {Leloudas}, {Levan}, {Mazzali}, \&
  {Pian}}]{Watson2019}
{Watson}, D., {Hansen}, C.~J., {Selsing}, J., {et~al.} 2019, \nat, 574, 497,
  \dodoi{10.1038/s41586-019-1676-3}

\bibitem[{{Waxman} {et~al.}(2018){Waxman}, {Ofek}, {Kushnir}, \&
  {Gal-Yam}}]{Waxman2018}
{Waxman}, E., {Ofek}, E.~O., {Kushnir}, D., \& {Gal-Yam}, A. 2018, \mnras, 481,
  3423, \dodoi{10.1093/mnras/sty2441}

\bibitem[{{Winteler} {et~al.}(2012){Winteler}, {K{\"a}ppeli}, {Perego},
  {Arcones}, {Vasset}, {Nishimura}, {Liebend{\"o}rfer}, \&
  {Thielemann}}]{2012ApJ...750L..22W}
{Winteler}, C., {K{\"a}ppeli}, R., {Perego}, A., {et~al.} 2012, \apjl, 750,
  L22, \dodoi{10.1088/2041-8205/750/1/L22}

\bibitem[{{Yang} {et~al.}(2015){Yang}, {Jin}, {Li}, {Covino}, {Zheng},
  {Hotokezaka}, {Fan}, {Piran}, \& {Wei}}]{Yang2015}
{Yang}, B., {Jin}, Z.-P., {Li}, X., {et~al.} 2015, Nature Communications, 6,
  7323, \dodoi{10.1038/ncomms8323}

\bibitem[{{Zevin} {et~al.}(2022){Zevin}, {Nugent}, {Adhikari}, {Fong}, {Holz},
  \& {Kelley}}]{2022ApJ...940L..18Z}
{Zevin}, M., {Nugent}, A.~E., {Adhikari}, S., {et~al.} 2022, \apjl, 940, L18,
  \dodoi{10.3847/2041-8213/ac91cd}

\bibitem[{{Zheng} \& {Ramirez-Ruiz}(2007)}]{2007ApJ...665.1220Z}
{Zheng}, Z., \& {Ramirez-Ruiz}, E. 2007, \apj, 665, 1220,
  \dodoi{10.1086/519544}

\end{thebibliography}

%\afterpage{\clearpage}

\appendix
\section{Mass and volume weighted distributions}
\label{app:distributions}
In Section~\ref{subsec:UMWabund} we discussed the evolution of the volume weighted mean and spread of [Fe/H], [Eu/H] and [Eu/Fe] in all the gas in the simulation volume, as well as, in the cold, dense gas in the disk. We showed that the global mean abundances are always higher than those of the cold, dense gas, and we argued that this is due to the larger volume filling factor of the gas in the wind region. We present here data to support this claim. 

In the top row of Figure~\ref{fig:abundances_w_slices} we reproduce Figure \ref{fig:UMW_abundances}, adding vertical lines to indicate three specific times in the simulation. These times are selected to more closely examine the structures of the volume and mass weighted abundance distributions. The subsequent three rows in the figure correspond to each of these time slices. In these rows the solid lines outline the volume weighted distributions (the color of the lines are the same as in the top row with darker color indicating cold, dense, disk gas while lighter color refers to the global gas values). We also note the mean value of these abundances with an arrow. The dashed lines show the equivalent distributions but weighted by gas mass rather than volume.

We first investigate the [Fe/H] distributions. The volume weighted distribution of all the gas shows a sharp peak at relatively high abundances, where the mean of the distribution lies. Furthermore, this distribution has a long tail on the lower abundance side, at all time slices considered. The volume weighted distribution of the cold, dense gas is much more symmetric and peaks at lower metallicity values. Thus, there are considerable differences between the two distributions when considering the volume weighted distributions. However, the mass weighted distributions are rather similar (dashed lines in Figure~\ref{fig:abundances_w_slices}). Hence, we conclude that the difference in mean abundance between the global and the cold, dense gas is driven by the higher abundance of the hot, diffuse gas in the disk winds being emphasized in the volume weighted distributions of the global gas. 

We now turn our attention to the [Eu/X] distributions and see many similar behaviors as seen in the [Fe/H] distributions. First, the volume weighted distribution of the global gas abundance is off-set at high metallicity when compared to all other distributions. Second, there is a nearly perfect overlap between the volume weighted distribution of the cold, dense gas and the mass weighted distributions of both the global and the cold, dense gas phases. Finally, the volume weighted global distribution features a tail at the low enrichment side. The distributions [Eu/X] are generally much broader than those of [Fe/H] indicating that the r-process elements are less well mixed in all gas phases than Fe is. As time increases in the simulation the offset between the mean of the global and the cold, dense disk gas slowly decreases. This is caused by gas mixing in the ISM. 

\begin{figure}
 \centering
 \includegraphics[height = 0.85\textheight]{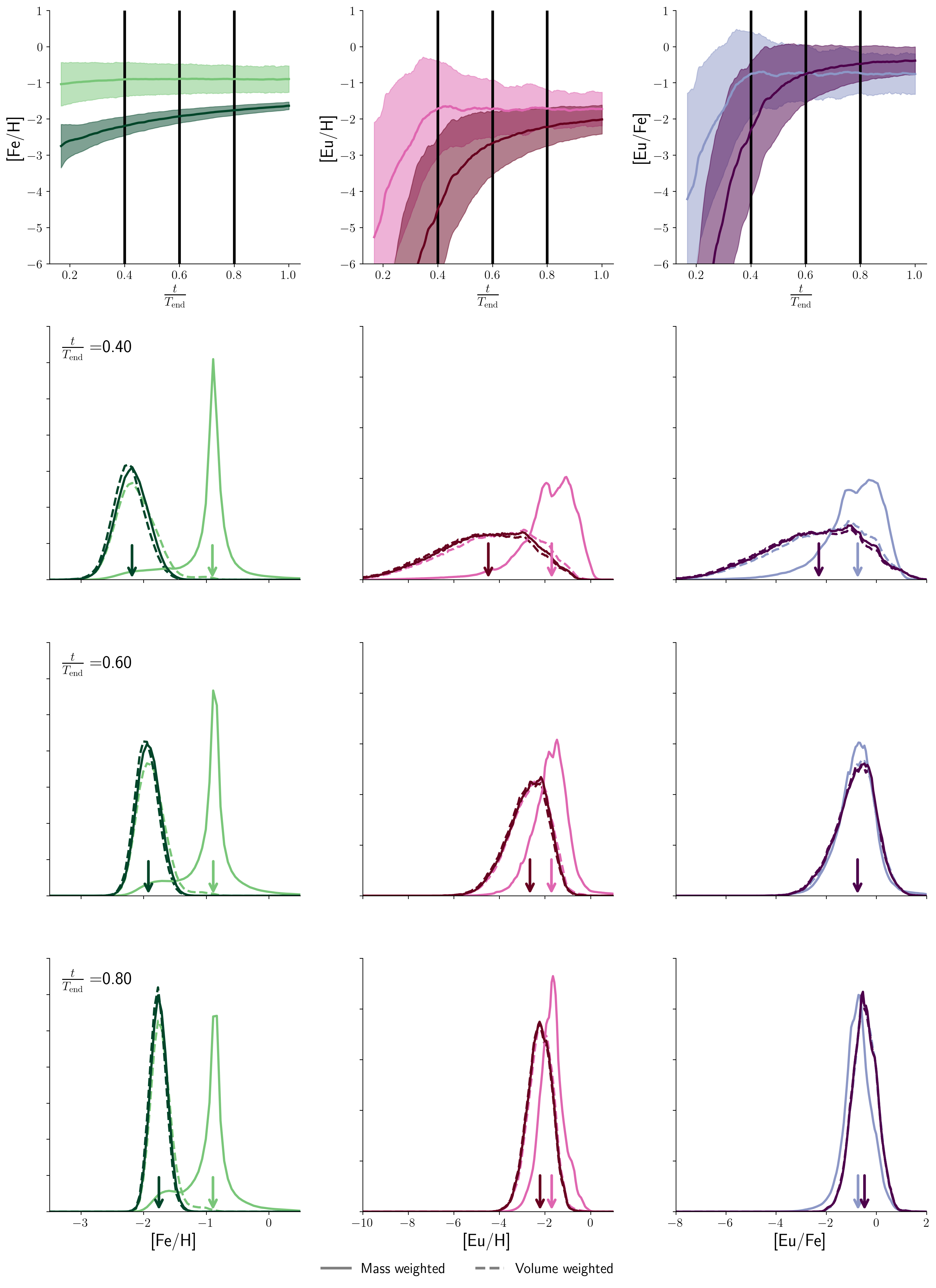}
 \caption{The abundance distributions at three distinct times in the metal mixing evolution of the MW progenitor with high SFR simulation. The top row is the same as Figure~\ref{fig:UMW_abundances}. The subsequent three rows show the distributions of [Fe/H], [Eu/H] and [Eu/Fe] at each of the three times indicated by vertical, black lines in the top row. The colors indicate the gas phase; lighter color for global values and darker for cold, dense disk gas. The solid lines show the volume weighted distributions while the dashed lines give the mass weighted distributions. The arrows show the mean value of the volume weighted distributions.}
 \label{fig:abundances_w_slices}
\end{figure}

\FloatBarrier
\section{Loading in all galaxy potentials}
\label{app:galaxy}
In Section~\ref{sec:potentials} we presented the loading factors of the total, iron and r-process mass in three galaxy potentials but we limited our attention to results for $\vert z \vert = 2 z_{\rm SNe}$. For completeness we present here the results for these loading factors at varying heights. In the middle row of Figure~\ref{fig:all_eta} we present an expanded version of Figure~\ref{fig:galaxy_mass_loading}. The top and bottom rows of the figure show the same corresponding loading factors at $\vert z \vert = 3 z_{\rm eff}$ and $\vert z \vert = {L \over 2} - dx$, respectively. Across all galaxy potentials, the loading factors are larger when measured closest to the disk and decrease as we approach the edge of the box. This effect is caused (partly) by circulation of material in the wind whereby some of the ejected material falls back. The drop in loading factor is strongest in the MW progenitor with high SFR and smallest in the satellite galaxy model indicating that material is more effeciently launched when residing in a weaker galaxy potential.

\begin{figure}
    \centering
    \includegraphics[height = 0.7\textheight]{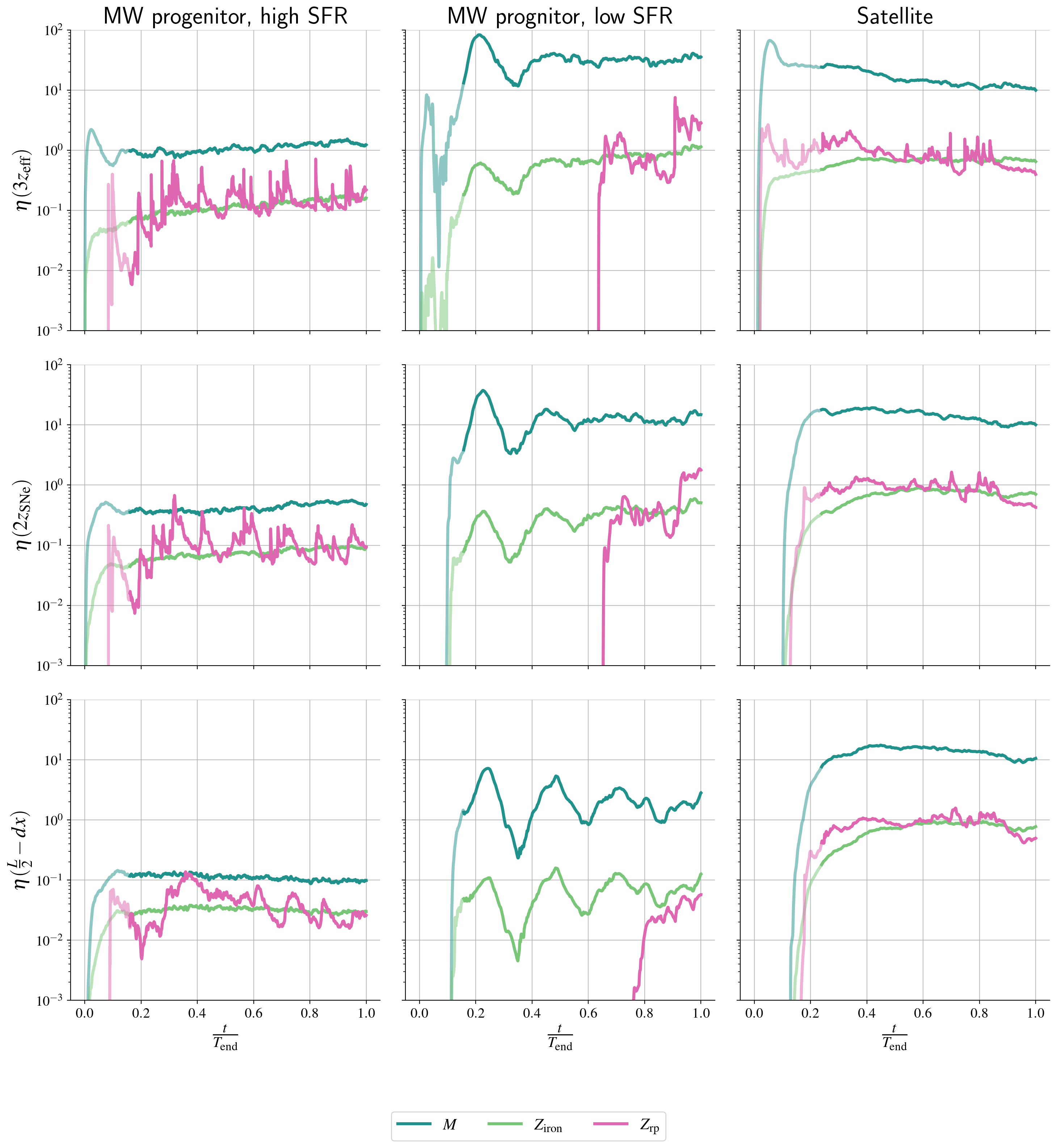}
    \caption{Loading factors for the total, iron and r-process mass (see legend) in each of the three galaxy patch models: MW progenitor with high SFR (left hand column), MW progenitor with low SFR (middle column) and satellite galaxy model (right hand column). Each row corresponds to mass outflow at increasing  distance from the disk mid-plane: $3 z_{\rm eff}$ (top row), $2 z_{\rm SNe}$(middle row) and ${L \over 2} - dx$ (bottom row). The middle row of this Figure is equivalent to Figure~\ref{fig:galaxy_mass_loading} although it has been extended to a slightly lower $\eta$ values.}
    \label{fig:all_eta}
\end{figure}

\FloatBarrier
\section{R-process production rate in the Milky Way}
\label{app:prod_rate}

\subsection{Light and heavy r-process elements}
We expand here on the presentation of our results in Sections~\ref{subsec:obs_match},~\ref{subsec:Hoto_comp} and~\ref{subsec:light_heavy}, and show how the abundances derived from various simulations compare with observations. Figure~\ref{fig:UMW_all_comp} shows the 50\% and 90\% contours of the distributions of [Eu/Fe] abundance as a function of the median [Fe/H] abundance of the cold, dense, disk gas. In the left (right) hand column we present the results when taking $A = 69$ ($A = 90$) as a lower limit for r-process elements. The r-process mass per event increases with each row, while the relative rate of r-process events is indicated by the color (see legend). We also include the 50\% and 90\% widths of the corresponding abundances of the metal-poor halo stars \citep{Roederer2014}, which are calculated in 0.23 dex wide [Fe/H] bins. 

In Section~\ref{subsec:obs_match} we showed the comparison between the best fitting r-process parameters and the observations. In Figure~\ref{fig:UMW_all_comp} we show how models with other parameter combinations fit the data. When choosing the best fit we take into account how well the simulations reproduce the mean of the observations, as well as, the abundance spread. This is done only for the range of metallicity over which the model is in steady state. We strongly disfavor models that predict a large number of stars with greater [Eu/Fe] abundances than those observed. 

In Section~\ref{subsec:light_heavy} we discussed the dependence on the constraints of $\dot{m}_{\rm rp}$ on the assumed range of r-process elements. Figure~\ref{fig:UMW_all_comp} shows how the distributions of [Eu/Fe] change as a function of the average [Fe/H] when we change $A_{\rm min}$. When an r-process pattern that includes only the heavy r-process elements ($A_{\rm min} = 90$) is assumed, the relative mass of Eu to all the r-process elements increases and thus, the distribution of [Eu/Fe] shifts by $+0.68$ dex. 

\begin{figure}
 \centering
 \includegraphics[height = 0.9 \textheight]{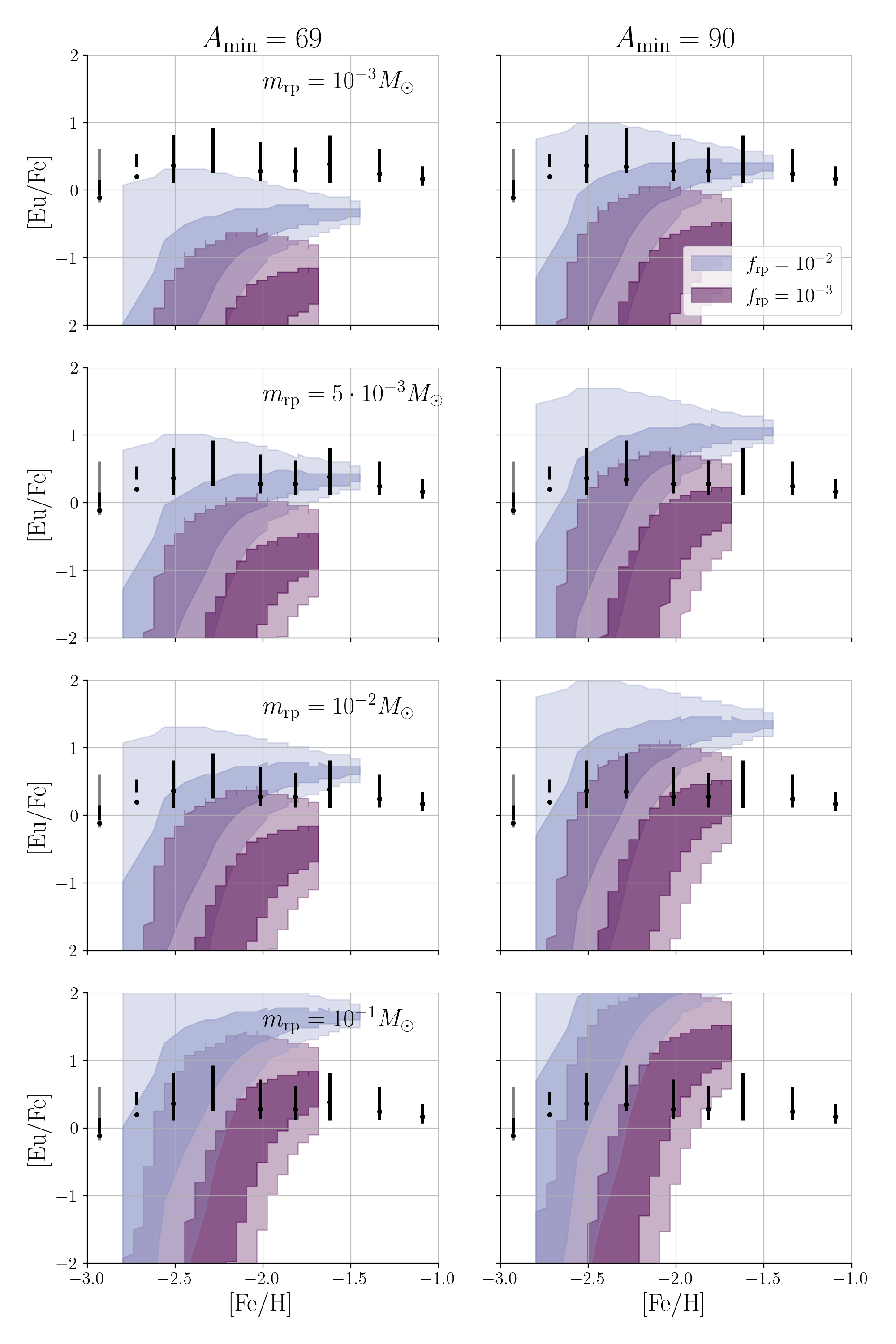}
 \caption{The simulated 50\% and 90 \% contours of the abundance distribution in the MW progenitor with high SFR when $m_{\rm rp}$ and $f_{\rm rp}$ are altered. The black and gray bars indicate the 50\% and 90\% spreads of the metal-poor MW halo stars \citep{Roederer2014} in bins of [Fe/H]. The left hand columns use the Eu mass fraction to all r-process elements, while the right hand uses the fraction relative only to the heavy r-process elements (defined here as $A> 90$).}
 \label{fig:UMW_all_comp}
\end{figure}

\subsection{Independence of $m_\mathrm{rp}$ and $f_{\rm rp}$}
In Figure \ref{fig:orthogonal} we explore the possible interdependence between $f_{\mathrm{rp}}$ and $m_{\rm rp}$. Along the top row of the figure the relative rate, $f_{\rm rp} = 10^{-2}$, is constant and the mass per event increases from left to right. Within each column the mass per event is constant but the relative rate decreases from the top to the bottom row such that the bottom rows have, $f_{\rm rp} = 10^{-3}$. The color of the shaded areas show the achieved average production rate of r-process elements per unit time in the simulation.

\begin{figure}
    \centering
    \includegraphics[width = \textwidth]{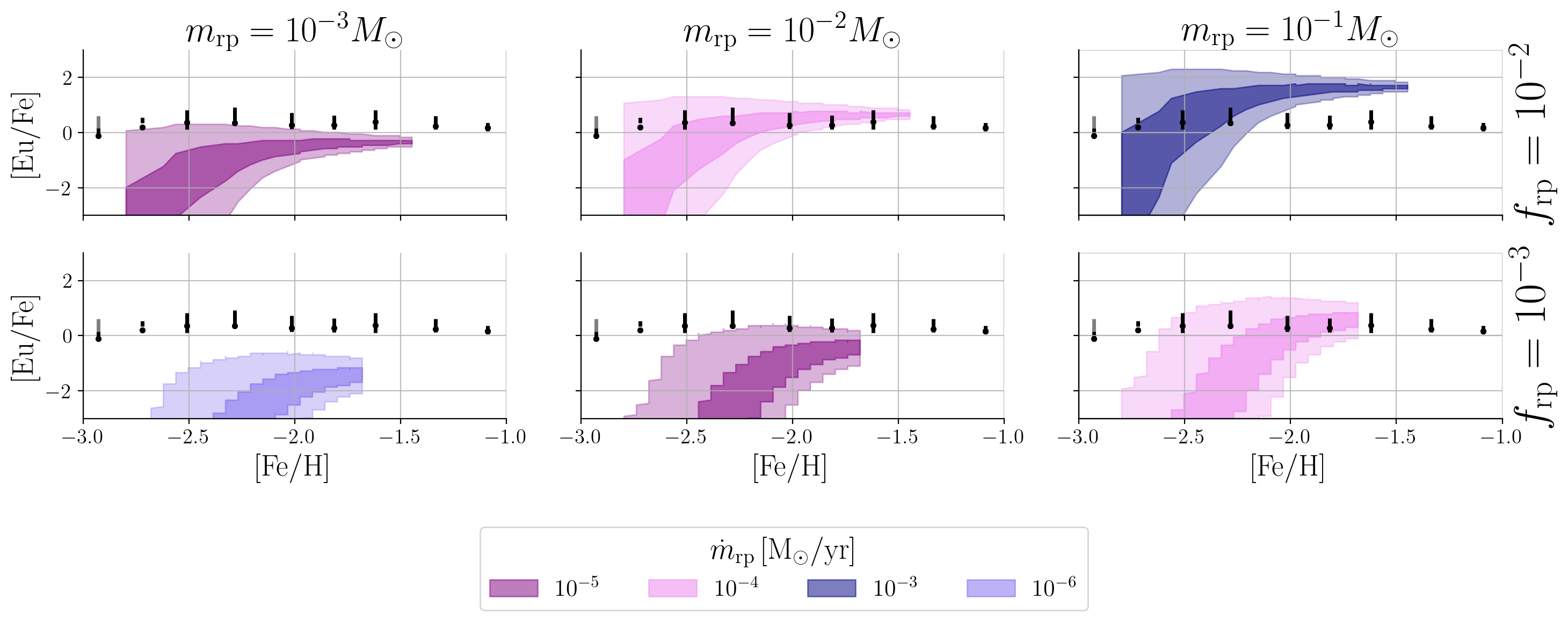}
    \caption{The same contours of the abundance distributions as the left hand column in Figure \ref{fig:UMW_all_comp} (omitting the $m_{\rm rp} = 5 \times 10^{-3} M_\odot$ case). Here the distributions are organized by the relative rate (constant along the rows) and mass per event (constant within a column) and color coded by the total r-process production rate.}
    \label{fig:orthogonal}
\end{figure}

\FloatBarrier

\end{document}